\documentclass[12pt]{article}
\usepackage{amssymb,amsfonts}
\usepackage{graphics,psboxit}
\usepackage{subfigure}
\usepackage{graphicx}
\usepackage{verbatim}
\usepackage{color}
\usepackage{dsfont}
\usepackage{hyperref}
\hypersetup{colorlinks}

\usepackage{enumerate}

\newenvironment{changemargin}[2]{%
\begin{list}{}{%
\setlength{\topsep}{0pt}%
\setlength{\leftmargin}{#1}%
\setlength{\rightmargin}{#2}%
\setlength{\listparindent}{\parindent}%
\setlength{\itemindent}{\parindent}%
\setlength{\parsep}{\parskip}%
}%
\item[]}{\end{list}}

\usepackage{amsmath}

\makeatletter
\def\adots{\mathinner{\mkern1mu\raise\p@
\vbox{\kern7\p@\hbox{.}}\mkern2mu
\raise4\p@\hbox{.}\mkern2mu\raise7\p@\hbox{.}\mkern1mu}}
\makeatother

\definecolor{darkred}{rgb}{1,0,0}
\definecolor{darkgreen}{rgb}{0,0.5,0}
\definecolor{darkblue}{rgb}{0,0,1}
\definecolor{orange}{rgb}{1,0.5,0}
\definecolor{green}{rgb}{0,1,0}
\definecolor{purple}{rgb}{.5,0,1}

\hypersetup{ colorlinks,
linkcolor=darkred,
filecolor=darkgreen,
urlcolor=darkblue,
citecolor=darkblue,
linktocpage=true }

\definecolor{markcolor}{rgb}{.25,0,1}

\definecolor{markcolor2}{rgb}{1,0,0}

\definecolor{markcolor3}{rgb}{0,1,0}

\def\be{\begin{equation}}
\def\ee{\end{equation}}
\def\ba{\begin{eqnarray}}
\def\ea{\end{eqnarray}}

\def\sgn{\textrm{sgn}}

\def\gg{\boldsymbol{\varphi}}
\def\ll{\textbf{b}}
\def\mm{\mathfrak{m}}
\def\nn{\mathfrak{n}}
\def\qf{\mathfrak{q}}

\def\hybrid{\topmargin -10pt    \oddsidemargin 0pt 
        \headheight 0pt \headsep 0pt
        \textwidth 16.5cm      
        \textheight 23cm       
        \marginparwidth .875in
        \parskip 5pt plus 1pt   \jot = 1.5ex}

\hybrid

\newcommand{\fr}[1]{\mathfrak{#1}}

\def\k{\kappa}
\def\r{\rho}
\def\a{\alpha}

\def\b{\beta}

\def\g{\gamma}

\def\d{\delta}

\def\m{\mu}
\def\n{\nu}

\def\l{\lambda}
\def\L{\Lambda}
\def\s{\sigma}

\def\no{\noindent}

\def\qq{\qquad}

\def\IR{\relax{\rm I\kern-.18em R}}

\csname @addtoreset\endcsname{equation}{section}

\begin{document}

\title{\textbf{Fermionic reflection matrices}}

\author{Nikos Karaiskos\footnote{\tt  nikolaos.karaiskos@itp.uni-hannover.de}}

\date{\normalsize{\textit{Institute for Theoretical Physics, Leibniz University Hannover,\\
Appelstrasse 2, 30167 Hannover, Germany}}}

\maketitle

\begin{abstract}
We consider the insertion of integrable boundaries for a class of supersymmetric quantum models. 
The generic conditions for constructing purely bosonic, purely fermionic or 
mixed type solutions of the graded 
reflection equation are extracted. Focusing on models associated with 
$\fr{gl}(\mm|\nn)$ or $\mathcal{U}_q(\fr{gl}(\mm|\nn))$ symmetry, we first
consider purely bosonic reflection matrices with special structures, for 
general values of $\mm, \nn$. These 
solutions provide the bosonic parts to construct fuller reflection matrices, 
containing fermionic degrees of freedom as well. 
\end{abstract}



%

\section{Introduction}
One dimensional quantum models can be treated very efficiently through the framework of Quantum Inverse
Scattering Method (QISM) \cite{QISM}. Despite their apparent simplicity, such models possess rich 
structures and even describe behaviors of certain materials (see for example \cite{Quantum_lab} and 
references therein). In this direction, the insertion of non-periodic boundary conditions is desirable
since finite-size effects may emerge. The systematic way of implementing boundaries which do not spoil
the integrability of the model was put forward in Sklyanin's seminal work \cite{reflection_early}. 

Since then, there has been a strong effort in constructing solutions of the 
reflection algebra. For 
diagonal reflection matrices a slight modification of the Algebraic Bethe Ansatz (ABA) \cite{reflection_early}
suffices for solving the model at hand. For nondiagonal reflection matrices a simple reference sate does 
not usually exist, so that the standard ABA cannot be applied. Various methods have been developed for such 
cases, especially for the XXZ open spin chain \cite{deVega:1993xi}, whose solution had been eluding for a long 
time \cite{XXZ_approaches}.

Although models of graded magnets appeared soon after the formulation of QISM
\cite{Kulish:1980ii}, supersymmetric extensions of models with integrable boundaries were 
considered much later in the literature. In spite of this delay, a great deal of graded 
models now exists, while many works have been devoted to the construction of 
reflection matrices associated with supersymmetric algebras \cite{deVega:1992zd}-\cite{Lima-Santos_slmn}. 
However, the large majority of these works inquires reflection matrices with zero fermionic  
degrees of freedom. On the other hand, the introduction of fermionic degrees 
of freedom is natural within the investigation of one dimensional electron
lattice models, such as the Hubbard and the supersymmetric t-J models \cite{EsslerKorepin}. 
This class of models constitutes a fruitful playground to explore 
condensed matter physics phenomena, particularly those including strongly
correlated electrons. Consideration of open boundaries in such models may lead 
to interesting physical behaviors, different than those of the ungraded ones. 
Furthermore, the grading has been proven to facilitate the search of 
exact solutions in at least two cases with nondiagonal boundary conditions:
the free fermion model \cite{Grab_Frahm} and the small polaron model 
\cite{Karaiskos:2013oca}, which may be regarded as graded versions of the 
open XX and XXZ spin chains respectively. In both cases supersymmetry lifts
the need of imposing any constraints on the boundary parameters, as 
happens for example in some approaches used to solve the open XXZ spin 
chain with nondiagonal boundaries. 

The present work aims to partially fill the aforementioned gap. Our survey
consists of obtaining results in two fronts: (i) the generic conditions 
for constructing reflection matrices with both bosonic and fermionic degrees of freedom
are extracted from the reflection algebra, for a large class of supersymmetric models and
(ii) focusing on the case of a $\fr{gl}(\mm|\nn)$ or $\mathcal{U}_q(\mathfrak{gl}(\mm|\nn))$ 
symmetry, we build \textit{special} solutions for generic values of $\mm$ and $\nn$ with both
bosonic and fermionic degrees of freedom switched on.

The structure of this paper is as follows: in Section 2 we begin with describing the basic objects
and setting up our conventions. A crucial notation is employed which enables to readily distinguish
between bosonic and fermionic degrees of freedom. Section 3 serves as a warm-up, where the conditions 
for generic, purely bosonic solutions of the reflection algebra are obtained and solutions with special
structures are rederived. New nondiagonal reflection matrices are also constructed for the $\fr{gl}(\mm|\nn)$
case. In Section 4 we consider reflection matrices with nonzero fermionic degrees 
of freedom for both the rational and the trigonometric cases. New reflection matrices are obtained, whose 
bosonic parts essentially correspond to those of Section 3. The matrices 
constructed in Section 4 are as far as we can tell novel. We will call
them \emph{fermionic reflection matrices} to distinguish them from 
reflection matrices with purely bosonic degrees of freedom.
We conclude with summing up our results and discussing possible future directions, while the Appendices
collect some bulky, but nevertheless useful material.

\section{The setting and conventions}
We assume that the vector spaces are $\mathbb{Z}_2$-graded \cite{Kac:1977em}, $V = V_0 \oplus V_1$ 
with dimensions $\dim V=\dim V_0 + \dim V_1= \mm + \nn $. 
Focusing on the case of a $\frak{gl}(\mm|\nn)$ symmetry or its
$q$-deformation, the following parities corresponding to the distinguished Dynkin diagram are 
assigned
\be
p(A) = \left\{ \begin{array}{ll}  0\,, & 1 \leq A \leq \mm \,,\\
			       1\,, & \mm + 1 \leq A \leq \mm + \nn  \, .
\end{array}\right.
\ee
The grading of the basis elements 
$(e_{AB})_{CD} = \d_{AC} \, \d_{BD}$ is considered to be 
\be
p(e_{AB}) = p(A) + p(B) \, , 
\ee
while the tensor product is graded by the following rule
\be
(e_{AB} \otimes e_{CD} ) ( e_{IJ} \otimes e_{KL}) = (-1)^{[p(C)+p(D)][p(I)+p(J)]} \, 
e_{AB}\, e_{IJ} \otimes e_{CD} \, e_{KL} \, .
\ee
The product of two basis elements is given by $e_{AB} \, e_{CD} = e_{AD} \, \d_{BC}$ and the graded
permutation operator is given by the expression
\be
\mathcal{P} = \sum_{A,B=1}^{\mm + \nn } (-1)^{p(B)} \, e_{AB} \otimes e_{BA} \, .
\label{graded_permu}
\ee
The $R$-matrix satisfies the graded Yang-Baxter equation \cite{Kulish:1980ii}
\be
R_{12}(\l) \, R_{13}(\l+\m) \, R_{23}(\m) \, = R_{23}(\m) \,R_{13}(\l+\m) \,R_{12}(\l) \, ,
\ee
and is assumed to have the following structure
\be
R(\l) = \sum_{A,B = 1}^{\mm + \nn} \Big(  f_{AB}(\l) \, e_{AA} \otimes e_{BB} \, 
+ g_{AB}(\l) \, e_{AB} \otimes e_{BA}\Big)\, .
\label{R_matrix}
\ee
Note that the conditions that we later derive are valid for all $R$-matrices with
this structure. However, in this paper we focus on the particular one associated with 
the $\mathcal{U}_q(\frak{gl}(\mm|\nn)$ algebra  with the following Boltzmann 
weights \cite{Perk:1981nb}
\be
\begin{split}
& f_{AB} = \sinh(\l)\, , \qq f_{AA}(\l) = \sinh(\l + i\eta -2i\eta \, p(A))\, , \\  
& g_{AA}(\l) = 0\, , \qquad g_{AB}(\l) = (-1)^{p(B)}\sinh(i\eta)\, e^{\sgn(B-A)\l}\, . 
\end{split}
\label{Boltzmann_weights_AB}
\ee
The rational limit $q=e^{i\eta} \to1 $ is readily obtained for the $\frak{gl}(\mm|\nn)$ symmetry 
so that the corresponding Boltzmann weights now read as \cite{Kulish:1980ii}
\be
\begin{split}
f_{AA}(\l) = \l + i \, (-1)^{p[A]} , \qquad f_{AB}(\l)  = \l\, , \cr
 g_{AA}(\l) = 0\, , \qquad g_{AB}(\l) = i \, (-1)^{(p[B])} \, .
\end{split}
\ee
In this case the $R$-matrix is just $R(\l) = \l \, \mathbb{I} + i \, \mathcal{P}$, with 
$\mathcal{P}$ given by (\ref{graded_permu}). The $R$-matrix fulfills the 
following properties \cite{Rmatrixprops}:
\be
\begin{split}
\textrm{Unitarity:} & \qquad R_{12}(\l) R_{21}(-\l) = \sinh(\l+i\eta) 
\sinh(-\l +i \eta)  \, \mathds{1} \cr 
\textrm{PT-symmetry:} & \qquad \mathcal{P}_{12} R_{12}(\l) \mathcal{P}_{12}
= R_{21}(\l) = R_{12}^{st_1\, st_2}(\l)    \cr
\textrm{Cross-unitarity:} & \qquad R_{12}^{st_1}(\l) M_1 
R_{12}^{st_2}(-\l-i \frak{\Delta} \eta ) M_1^{-1} = \sinh(\l) 
\sinh(-\l-i\frak{\Delta}\eta) \, \mathds{1} \, ,
\end{split}
\label{Rmatrix_properties}
\ee
with $\frak{\Delta} \equiv \mm - \nn$ and the supertrace is performed via the operation
\be
A^{st} = \sum_{B,C=1}^{\mm+\nn} A_{BC} \, e_{BC}^{st} =  \sum_{B,C=1}^{\mm+\nn} 
(-1)^{p(B)+p(B)p(C)}A_{BC} \, e_{CB} \, .
\ee
The so-called crossing matrix introduced above, $M$, is an 
$(\mm+\nn) \times (\mm+\nn)$ diagonal matrix defined by
\be
M = \sum_{A=1}^{\mm+\nn} M_{A} \, e_{AA} \, , \qquad 
M_A = \left\{ 
\begin{array}{cc}
 e^{2i\eta(1-A)}\, , & 1 \leq A \leq \mm \,  \cr
 -e^{2i\eta(A-2\mm)}\, , & \mm < A \leq \mm + \nn \, ,
\end{array}
\right.
\ee
and is a symmetry of the $R$-matrix
\be
[R_{12}(u) \, , \, M \otimes M ] =0 \, . 
\ee
Alongside the $R$-matrix, the reflection matrices satisfy the graded reflection algebra \cite{reflection_early, Bracken}
\be
R_{12}(\l-\m)\,  K_1^-(\l) \, R_{21}(\l+\m) \, K_2^-(\m) = K_2^-(\m) \, 
R_{12}(\l+\m) \, K_1^-(\l) \, R_{21}(\l-\m) \, .
\label{graded_reflection_algebra}
\ee
Our aim is to solve the reflection equation by considering the most general expression for the  reflection matrix 
\be
K^-(\l) = \sum_{A,B = 1}^{\mm + \nn}  h_{AB}(\l) \, e_{AB} \, ,
\label{K_general}
\ee
where the unknown functions $h_{AB}(\l)$ to be computed are either commuting or anticommuting functions of the 
spectral parameter, their nature being determined by the grading of the corresponding indices. A second reflection equation provides the reflection
matrices for the other boundary \cite{Bracken}
\be 
\begin{split}
& R_{12}(-\l + \m) \, K_1^+(\l) \, M_1^{-1} \, R_{21}(-\l-\m-2i \frak{\Delta} \eta) \, M_1 \, K_2^+(\m) \cr
& ~~ = K_2^+(\m) \, M_1 \, R_{12}(-\l-\m-2i\frak{\Delta}\eta) \, M_1^{-1}
\, K_1^+(\l) \, R_{21}(-\l + \m) \, . 
\end{split}
\label{second_refl_alg}
\ee
By virtue of the crossing unitarity property (\ref{Rmatrix_properties}), there
exists an isomorphism between the graded reflection algebras 
(\ref{graded_reflection_algebra}) and (\ref{second_refl_alg}), which implies 
that \cite{MeziNepo}
\be
K^+(\l) = K^-(-\l - i\frak{\Delta}\eta) \, M \, .
\ee
Since the construction of the reflection matrices $K^-(\l)$ provides also the 
solutions $K^+(\l)$ through this isomorphism, we will henceforth 
restrict ourselves to the derivation of the first ones. We will also relax the
notation by dropping the minus sign and simply 
denote $K(\l) \equiv K^-(\l)$. 

At this point, we employ an important notation: we split the capital indices 
$\{A\}  \in \big[1, \mm + \nn  \big]$ 
into small ones. The bosonic (fermionic) degrees of freedom will be represented by indices denoted by 
small latin (greek) letters
\[
 \{ \ll \} \in \big[1, \mm\big] \qquad \textrm{and} \qquad \{ \gg \} \in \big[\mm+1, \mm+\nn\big] \, . 
\]
As a consequence, the various gradings appearing are  abstractly computed: 
\[
 p(\ll) = 0 \, \qquad \textrm{and} \qquad p(\gg) = 1 \, .
\]
In this spirit, the $K$-matrix (\ref{K_general}) will be given by 
\be
K(\l) = \sum_{\substack{a=1\\b=1}}^\mm  h_{ab}(\l) \, e_{ab} + 
\sum_{a=1}^\mm \sum_{\s=\mm+1}^{\mm + \nn } \chi_{a\s}(\l) \, e_{a\s} 
+\sum_{\r=\mm+1}^{\mm + \nn}  \sum_{b=1}^\mm \chi_{\r b}(\l) \, e_{\r b}  + 
\sum_{\substack{\r=\mm+1 \\ \s=\mm+1}}^{\mm + \nn}  h_{\r\s}(\l) \, e_{\r \s} \, ,
\ee
where the functions containing Grassmann variables are now denoted by $\chi(\l)$, whereas $h(\l)$ are purely 
bosonic functions. A similar expansion is to be understood for the $R$-matrix, where the Boltzmann weights 
(\ref{Boltzmann_weights_AB}) now read
\be
\begin{split}
& f_{aa}(\l) = \sinh(\l +i\eta) \, , \qquad f_{\r\r}(\l) = \sinh(\l - i\eta) \, , \\
& f_{ab}(\l) = f_{\r b}(\l) = f_{a\s}(\l) = f_{\r\s}(\l) = \sinh(\l)\, , \\
& g_{ab}(\l) = \sinh(i\eta) e^{\sgn(b-a)\l} \, , \qquad g_{\r\s}(\l) = -\sinh(i\eta) e^{\sgn(\s-\r)\l} \, , \\
& g_{\r b}(\l) = \sinh(i\eta) e^{-\l} \, , \qquad g_{a\s}(\l) = - \sinh(i\eta) e^\l \, , 
\qquad  g_{aa}(\l) = g_{\r\r}(\l) = 0 \, .
\end{split}
\label{Boltzmann_weights}
\ee
Substituting the expressions (\ref{R_matrix})  and (\ref{K_general}) into the graded
reflection algebra (\ref{graded_reflection_algebra}) produces $8^4$ terms at each side of 
the equation. However, the 
vast majority of them vanishes identically since they  contain terms 
proportional to $\d_{\ll \gg}$, which is 
zero by definition. Eventually, only $8^2$ nonzero terms survive, which may 
further be grouped with respect to the basis elements they contain. After appropriate relabellings, 
one is left with the relations which determine the unknown functions $h(\l)$ and 
$\chi(\l)$. There exist three types of conditions: those containing only the bosonic functions 
$h(\l)$, those containing only the fermionic functions $\chi(\l)$ and those involving both of them.
Before proceeding to the complete case with both bosonic and fermionic degrees
of freedom switched on, we first consider purely bosonic 
reflection matrices for both the rational and the trigonometric cases.

\section{Purely bosonic reflection matrices}
Setting $\chi(\l)=0$ further reduces the number of relations emerging from
the reflection equation, which can be gathered into six different sets, depending on
the grading of the basis elements $e_{AB} \otimes e_{CD}$. The bosonic (fermionic)
nature of the indices $A,B,C,D$ will be denoted by $\ll$ ($\gg$). It should be
noted that in the following relations no summation takes place, unless explicitly 
stated. For the sake of presentation, we also adopt the following conventions to 
represent the arguments of the various functions:
\[
F_{AB}(\l) \equiv F_{AB}^\l \, , \qquad \textrm{and} \qquad F_{AB}(\l\pm\m) \equiv F_{AB}^\pm \, . 
\]
The first two constraints emerging from the reflection algebra read  as
\begin{subequations}
\begin{flalign}
 \gg\gg\ll\ll: \qq &
\big(f_{k\r}^+ \, f_{\g k}^- - f_{j\r}^- \, f_{\g j}^+\big) h_{kj}^\m \, h_{\g\r}^\l
 + \d_{\g\r} \sum_d g_{\r d}^+ \big(g_{\r k}^- \, h_{d j}^\m \, h_{kd}^\l 
 - g_{\r j}^- \, h_{dj}^\l \, h_{kd}^\m\big) = 0
 \label{constr1} & \\
 \ll\ll\gg\gg: \qq & 
 \big(f_{\s j}^+ \, f_{c \s}^- - f_{\b j}^- \, f_{c \b}^+\big) h_{\s \b}^\m \, h_{c j}^\l
 + \d_{c j} \sum_\d g_{j \d}^+ \big(g_{j \s}^- \, h_{\d \b}^\m \, h_{\s\d}^\l 
 - g_{j \b}^- \, h_{\d \b}^\l \, h_{\s\d}^\m\big) = 0 \, . &
 \label{constr2}
\end{flalign}
\end{subequations}
For the Boltzmann weights (\ref{Boltzmann_weights}) in particular, these two conditions  simplify to 
\begin{subequations}
\begin{flalign}
& \gg\gg\ll\ll: \qq 
  \sum_d  \big( h_{d j}^\m \, h_{kd}^\l -  h_{dj}^\l \, h_{kd}^\m\big) = 0  & 
  \label{cons_simple1} \\
& \ll\ll\gg\gg: \qq  
   \sum_\d  \big( h_{\d \b}^\m \, h_{\s\d}^\l  -  h_{\d \b}^\l \, h_{\s\d}^\m\big) = 0  \, . &
\label{cons_simple2}
\end{flalign}
\end{subequations}
Proceeding to the other terms, one finds another pair of conditions, which read as
\begin{subequations}
\begin{flalign}
& \gg\ll\ll\gg: & \nonumber \\
& \d_{kj} \, f_{\g j}^- \sum_\d g_{j\d}^+ \, h_{\g\d}^\l \, h_{\d\r}^\m  
+ \d_{\g\r} \, f_{\r j}^- \sum_d g_{\r d}^+ \, h_{dj}^\l \, h_{kd}^\m 
 -f_{\g j}^+ \big( g_{j\r}^- \, h_{\g\r}^\l \, h_{kj}^\m 
+ g_{\g k}^- \, h_{kj}^\l \, h_{\g\r}^\m \big) =0 &
 \\ 
 & \ll\gg\gg\ll:  & \nonumber \\
 & \d_{\s\g} \, f_{c\s}^- \sum_d g_{\s d}^+ \, h_{cd}^\l \, h_{dj}^\m
+ \d_{cj} \, f_{j\s}^- \sum_\d g_{j\d}^+ \, h_{\d\s}^\l \, h_{\g\d}^\m 
- f_{c\s}^+ \big(g_{c\g}^- \, h_{\g\s}^\l \, h_{cj}^\m  
+ g_{\s j}^- \, h_{cj}^\l \, h_{\g\s}^\m \big) =0 \, . &
\label{boson_middle}
\end{flalign}
\end{subequations}
Equating the latter two expressions and after some index relabelling, one is 
consistently lead to (\ref{cons_simple1}), (\ref{cons_simple2}). There are two more 
sets of relations, more complicated than the previous ones, which read
\begin{subequations}
\begin{flalign}
\ll\ll\ll\ll: \qquad & \big( f_{ck}^- \, f_{kb}^+ - f_{cj}^+ \, f_{kb}^- \big) h_{cb}^\l  \, h_{kj}^\m  
+ f_{cb}^+ \big(g_{ck}^- \, h_{kb}^\l  \, h_{cj}^\m - g_{bj}^-  \, h_{cj}^\l
h_{kb}^\m \big) 
+ \d_{kb} \, f_{cb}^-  \sum_d g_{bd}^+ \, h_{cd}^\l  \, h_{dj}^\m & \nonumber \\
&   - \d_{cj} \,
f_{jb}^-  \sum_d g_{jd}^+  \, h_{db}^\l h_{kd}^\m  
+ \d_{cb} \sum_d g_{bd}^+ \big( g_{bk}^- \, h_{kd}^\l  \, h_{dj}^\m
- g_{bj}^- \, h_{dj}^\l  \, h_{kd}^\m \big) =0 
\label{last_bosonic1}\\
\gg\gg\gg\gg: \qquad  & \big(f_{\g\k}^- \, f_{\k\s}^+  -  f_{\g\r}^+ \, f_{\r\s}^-\big) h_{\g\s}^\l  \, h_{\k\r}^\m  
+ f_{\g\s}^+ \big(g_{\g\k}^- \, h_{\g\r}^\m \, h_{\k\s}^\l
- g_{\s\r}^-  \, h_{\g\r}^\l \, h_{\k\s}^\m \big) 
+ \d_{\k\s} \, f_{\g\s}^-   \sum_\d g_{\s\d}^+ \, h_{\g\d}^\l \, h_{\d\r}^\m & \nonumber \\
&  - \d_{\g\r} \,  f_{\r\s}^-  \sum_\d g_{\r\d}^+ \, h_{\d\s}^\l \, h_{\k\d}^\m 
 + \d_{\g\s} \sum_\d g_{\s\d}^+  \big(g_{\s\k}^- \, h_{\d\r}^\m \, h_{\k\d}^\l  - 
    g_{\s\r}^- \, h_{\d\r}^\l \, h_{\k\d}^\m\big) =0 \, .&
\label{last_bosonic2}
\end{flalign}
\end{subequations}
These six relations are in principle sufficient to provide all $c$-number, purely 
bosonic solutions to the reflection equation for every given $R$-matrix with the structure 
(\ref{R_matrix}). In practice, however, a derivation of generic solutions is a hard task, 
so that one usually restricts to diagonal reflection matrices, or
nondiagonal ones with special structures. In the present context, consideration of such 
solutions below 
has a two-fold importance, since (i) it serves as a validity check of our approach
while obtaining previous results and (ii) it provides us with (new) bosonic solutions that 
will be later used to construct reflection matrices with nonzero fermionic degrees of 
freedom.

\subsection{Special solutions for the rational case}
\subsubsection*{The trivial boundary}
It is easily checked that by setting $h_{AB}(\l) = w(\l) \, \d_{AB}$, all conditions 
(\ref{constr1})-(\ref{last_bosonic2}) are satisfied for the case of a trivial boundary.

\subsubsection*{Diagonal solutions}
The simplest non trivial solutions are obtained by introducing an index dependence,
i.e. by setting $h_{AB}(\l) = w_A(\l) \, \d_{AB}$.  Diagonal reflection matrices have been classified in
the past for various algebras \cite{deVega:1992zd, Arnaudon:2003gj}. For 
the $\mathfrak{gl}(\mm|\nn)$ case it is 
known that the diagonal reflection matrices are given by \cite{Arnaudon:2003gj, Ragoucy:2007kg}
\be
K(\l) = \textrm{diag} \big( \underbrace{c_0 - \l , \cdots , c_0 - \l}_{\mathfrak{q}_1}, \,
\underbrace{c_0 + \l , \cdots , c_0 + \l}_{\mathfrak{q}_2- \qf_1}, \, 
\underbrace{c_0 - \l , \cdots , c_0 - \l}_{\mm+\nn-\qf_2} \big) \, ,
\ee
with $c_0$ being a free complex parameter and $0\leq \mathfrak{q}_1 \leq \qf_2 \leq \mm+\nn$.

\subsubsection*{Nondiagonal solutions}
Apart from purely diagonal entries of the reflection matrix, one may obtain 
nondiagonal ones as well, for generic values of $\mm,\nn$. In these solutions, the nonzero entries of 
the reflection matrix lie solely along the principal and the secondary diagonals. Define the conjugate 
index
\be
 \bar{a} = \mm + 1 - a \qquad \textrm{and} \qquad 
 \bar{\r} = 2\mm + \nn + 1 - \r \, ,
 \label{conjugate_index}
\ee
so that a generic element of the $K$-matrix would have the following expression 
\be
h_{AB}(\l) = w_A(\l) \, \d_{AB} + z_A(\l) \, \d_{B\bar{A}}\,.
\label{nond_struct_h}
\ee
Substituting into the conditions (\ref{constr1})-(\ref{last_bosonic2}) and inspired by \cite{Galleas:2007zz},
we are able to obtain nondiagonal solutions for the distinguished Dynkin diagram that we consider in the present work.  
Note that the nonzero nondiagonal entries have either purely bosonic, or purely fermionic 
indices and that all solutions have five free complex parameters $c_0, \ldots c_4$. Defining 
\be
\begin{split}
& \Xi_{\ll\ll}^\pm(\l) \equiv c_0 \pm \frac{\l (c_1c_2 + c_3c_4 )}{c_3 c_4 - c_1c_2} \, ,
 \qquad \Xi_{\gg\gg}^\pm(\l) \equiv c_0 \pm \l \, , \cr
 &  \Xi_{14}(\l) \equiv \frac{2\l \, c_1 c_4}{c_1 c_2 - c_3 c_4} \, , \qquad
  \Xi_{23}(\l) \equiv \frac{2\l \, c_2 c_3}{c_3 c_4 - c_1c_2} \, , 
\end{split}
 \ee
we have found the following families of solutions ($[x]$ denotes the integer part of $x$): 

\textbf{I$\ll$.} Solutions with bosonic nondiagonal indices for  $1\leq L \leq \left[\frac{\mm}{2}\right]$:
\be
\begin{split}
& h_{\bar{\r}\r}(\l)=h_{\r\bar{\r}}(\l) = 0 \,,  \qquad 
h_{\r\r}(\l) = h_{\bar{\r}\bar{\r}}(\l) = \Xi_{\gg\gg}^-(\l) \, , \\
& h_{jj} (\l)= \Xi_{\ll\ll}^+(\l)  , \quad h_{\bar{j}\bar{j}}(\l) = \Xi_{\ll\ll}^-(\l), 
\quad  h_{j\bar{j}}(\l) = \Xi_{14}(\l), \quad h_{\bar{j}j}(\l) = \Xi_{23}(\l) \, , \quad 1\leq j\leq L \\
& h_{jj}(\l) = h_{\bar{j}\bar{j}}(\l) =  \Xi_{\gg\gg}^-(\l) \, , \qquad 
h_{j\bar{j}} (\l)= h_{\bar{j}j} (\l)= 0 \, , \qquad L < j\leq \left[\frac{\mm}{2}\right]  \\
& h_{jj}(\l) = h_{\bar{j}\bar{j}}(\l) = \Xi_{\gg\gg}^-(\l)
\, , \qquad j = \bar{j} = \frac{\mm+1}{2} \, , 
\qquad \textrm{if} ~~ \mm ~~ \textrm{odd} \, .
\end{split}
\ee

\textbf{I$\gg$.}  Solutions with fermionic nondiagonal indices for $\mm+1 \leq \L \leq \mm + 
\left[\frac{\nn}{2}\right]$:
 \be
 \begin{split}
  & h_{j\bar{j}} (\l)= h_{\bar{j}j} (\l)= 0\, , \qquad 
  h_{jj}(\l) = h_{\bar{j}\bar{j}}(\l) = \Xi_{\gg\gg}^-(\l) \, , \\
  & h_{\r\r} (\l) = \Xi_{\ll\ll}^+(\l) , \quad h_{\bar{\r}\bar{\r}}(\l) = \Xi_{\ll\ll}^-(\l)  , \quad
  h_{\r\bar{\r}}(\l) = \Xi_{14}(\l) \, , \quad 
  h_{\bar{\r}\r}(\l) = \Xi_{23}(\l) , \quad \mm +1 \leq \r \leq \L\,, \\
  & h_{\r\r}(\l)= h_{\bar{\r}\bar{\r}}(\l) = \Xi_{\gg\gg}^-(\l) \, , \quad 
  h_{\bar{\r}\r}(\l)=h_{\r\bar{\r}}(\l) = 0 \, , \qquad \L < \r \leq \mm+\left[\frac{\nn}{2}\right] \\
  & h_{\r\r}(\l) = h_{\bar{\r}\bar{\r}}(\l) = \Xi_{\gg\gg}^-(\l) \, , \qquad 
  \r = \bar{\r} = \mm + \frac{\nn+1}{2} \, , \qquad \textrm{if} ~~ \nn ~~ \textrm{odd} \, .
 \end{split}
 \ee

It is possible to construct additional solutions with nondiagonal entries, by defining 
appropriate generalized conjugate indices, $a_\ell^\pm, \r_\xi^\pm$, as explained 
in Appendix A. Running over these generalized conjugate indices amounts to 
spanning minor anti-diagonals below and 
above the secondary diagonal of the bosonic submatrices. Assuming that the 
reflection matrix under determination has nonzero entries only along the diagonal 
and one such minor anti-diagonal, 
that is of the form (\ref{nond_struct_h}) again, then the above results hold for these cases 
as well, with $(L, \L, j, \r )$ now taking values into smaller intervals. We have found 
the following four distinct families of solutions, categorized with respect to $L,\L$:

\textbf{II$\ll$.} Solutions with bosonic nondiagonal indices for  $1\leq L \leq \frac{\mm}{2} - 
\left[ \frac{\ell}{2}\right] $:
\be
\begin{split}
& h_{\r_\xi^\pm\r}(\l)=h_{\r\r_\xi^\pm}(\l) = 0 \,,  \qquad 
h_{\r\r}(\l) = h_{\r_\xi^\pm\r_\xi^\pm}(\l) = \Xi_{\gg\gg}^-(\l) \, , \\
& h_{jj} (\l)= \Xi_{\ll\ll}^+(\l)  , \quad h_{j_\ell^-j_\ell^-}(\l) = \Xi_{\ll\ll}^-(\l), 
\quad  h_{jj_\ell^-}(\l) = \Xi_{14}(\l) \, , \quad h_{jj}(\l) = \Xi_{23}(\l) \, , \quad 1\leq j\leq L  \\
& h_{jj}(\l) = h_{j_\ell^-j_\ell^-}(\l) =  \Xi_{\gg\gg}^-(\l) \, , \quad 
h_{jj_\ell^-} (\l)= h_{j_\ell^-j} (\l)= 0 \, , ~~~ L < j \leq \frac{\mm}{2} - 
\left[ \frac{\ell}{2}\right] \\
& h_{jj} (\l) = h_{j_\ell^-j_\ell^-}(\l)  = \Xi_{\gg\gg}^-(\l) \, , \qquad 
j=j_\ell^- = \frac{\mm+1}{2}
-\left[\frac{\ell}{2}\right] \qquad \textrm{if} ~~ \mm ~~\textrm{odd} \\
&  h_{jj}(\l) = \Xi_{\gg\gg}^-(\l) \, , \qquad \mm+1 - \ell \leq j \leq \mm \, .
\end{split}
\ee

\textbf{III$\ll$.} Solutions with bosonic nondiagonal indices for  $\ell +1 \leq L \leq \frac{\mm}{2} + 
\left[\frac{\ell}{2}\right]$:
\be
\begin{split}
& h_{\r_\xi^\pm\r}(\l)=h_{\r\r_\xi^\pm}(\l) = 0 \,,  \qquad 
h_{\r\r}(\l) = h_{\r_\xi^\pm\r_\xi^\pm}(\l) = \Xi_{\gg\gg}^-(\l) \, , \\
& h_{jj}(\l) =  \Xi_{\gg\gg}^-(\l) \, , \qq 1 \leq j \leq \ell \\
& h_{jj} (\l)= \Xi_{\ll\ll}^+(\l)  , \quad h_{j_\ell^+j_\ell^+}(\l) = \Xi_{\ll\ll}^-(\l), 
\quad  h_{jj_\ell^+}(\l) = \Xi_{14}(\l) \, , \quad h_{j_\ell^+j}(\l) = \Xi_{23}(\l) \, , \quad \ell  < j\leq L \\
& h_{jj}(\l) = h_{j_\ell^+j_\ell^+}(\l) =  \Xi_{\gg\gg}^-(\l) \, , \quad
h_{jj_\ell^+} (\l)= h_{j_\ell^+j} (\l)= 0 \, , \quad L < j \leq  
\frac{\mm}{2} +  \left[\frac{\ell}{2}\right] \\
& h_{jj}(\l) = h_{j_\ell^+j_\ell^+}(\l) =  \Xi_{\gg\gg}^-(\l) \, , \qquad 
j=j_\ell^+ = \frac{\mm+1}{2} + \left[\frac{\ell}{2}\right] \qquad
\textrm{if} ~~ \mm ~~ \textrm{odd}
\end{split}
\ee

\textbf{II$\gg$.} Solutions with fermionic nondiagonal indices for $\mm+1 \leq \L \leq \mm + \frac{\nn}{2} 
- \left[\frac{\xi}{2}\right]$:
\be
 \begin{split}
  & h_{jj_\ell^\pm} (\l)= h_{j_\ell^\pm j} (\l)= 0\, , \qquad 
  h_{jj}(\l) = h_{j_\ell^\pm j_\ell^\pm}(\l) = \Xi_{\gg\gg}^-(\l) \, , \\
  & h_{\r\r} (\l)= \Xi_{\ll\ll}^+(\l) , \quad h_{\r_\xi^- \r_\xi^-}(\l) = \Xi_{\ll\ll}^-(\l)  , \quad
  h_{\r\r_\xi^-}(\l) = \Xi_{14}(\l) \, , \quad h_{\r_\xi^-\r}(\l) = \Xi_{23}(\l), \quad \mm +1 \leq \r \leq \L \\
  & h_{\r\r}(\l)= h_{\r_\xi^-\r_\xi^-}(\l) = \Xi_{\gg\gg}^-(\l) \, , \quad 
  h_{\r_\xi^-\r}(\l)=h_{\r\r_\xi^-}(\l) = 0 \, , \quad \L < \r \leq \mm+\frac{\nn}{2} - \left[\frac{\xi}{2}\right] \\
  & h_{\r\r}(\l)= h_{\r_\xi^-\r_\xi^-}(\l) = \Xi_{\gg\gg}^-(\l) \, , \quad
  \r = \r_\xi^- = \mm + \frac{\nn+1}{2} - \left[\frac{\xi}{2}\right] \quad 
  \textrm{if}~~ \nn ~~ \textrm{odd} \\
  & h_{\r\r}(\l) = \Xi_{\gg\gg}^-(\l)  \, , \qquad  \mm + \nn + 1 - \xi \leq \r \leq \mm + \nn
 \end{split}
 \ee

 \textbf{III$\gg$.} Solutions with fermionic nondiagonal indices for $\mm+1 \leq \L \leq \mm + \frac{\nn}{2} 
+ \left[\frac{\xi}{2}\right]$
\be
 \begin{split}
  & h_{jj_\ell^\pm} (\l)= h_{j_\ell^\pm j} (\l)= 0\, , \qquad 
  h_{jj}(\l) = h_{j_\ell^\pm j_\ell^\pm}(\l) = \Xi_{\gg\gg}^-(\l) \, , \\
  & h_{\r\r}(\l) = \Xi_{\gg\gg}^-(\l) \, , \qquad \mm +1 \leq \r \leq \mm + \xi \\
  & h_{\r\r} (\l)=\Xi_{\ll\ll}^+(\l) , \quad h_{\r_\xi^+\r_\xi^+}(\l) = \Xi_{\ll\ll}^-(\l)  , \quad
  h_{\r\r_\xi^+}(\l) = \Xi_{14}(\l) \, , \quad h_{\r_\xi^+\r}(\l) = \Xi_{23}(\l) , \quad \mm + \xi  < \r \leq \L\\
  & h_{\r\r}(\l)= h_{\r_\xi^+\r_\xi^+}(\l) = \Xi_{\gg\gg}^-(\l) \, , \quad 
  h_{\r\r_\xi^+}(\l)=h_{\r_\xi^+\r}(\l) = 0 \, , \quad \L < \r \leq \mm+\frac{\nn}{2}  + \left[\frac{\xi}{2}\right] \\
  & h_{\r\r}(\l)= h_{\r_\xi^+\r_\xi^+}(\l) = \Xi_{\gg\gg}^-(\l) \, , \quad 
  \r = \r^+ = \m + \frac{\nn+1}{2} + \left[\frac{\xi}{2}\right] \quad 
  \textrm{if} ~~ \nn ~~ \textrm{odd}
 \end{split}
 \ee
 The nondiagonal solutions $\textbf{I}\ll - \textbf{III}\gg$ will be used in Section 4 as the 
 bosonic parts of fuller reflection matrices containing fermionic degrees of freedom as well.

\subsection{Special solutions for the trigonometric case}

\subsubsection*{Diagonal solutions}
The $\mathcal{U}_q(\mathfrak{gl}(\mm|\nn))$ case is more restricted than the rational one.  
The corresponding diagonal reflection matrices are now given by the expressions \cite{Arnaudon:2003gj}
\be
K(\l) = \textrm{diag} \big( \underbrace{c_0\, e^{\l} , \cdots , c_0 \, e^{\l}}_{\mathfrak{q}}, \,
\underbrace{c_0 \, e^{-\l}, \cdots , c_0 \, e^{-\l}}_{\mm + \nn - \mathfrak{q}} \big) \, ,
\ee
with $c_0$ being a free complex parameter and $0\leq \mathfrak{q} \leq \mm+\nn$.

\subsubsection*{Nondiagonal solutions}
Similarly to the rational case, there also exist nondiagonal solutions for the 
$\mathcal{U}_q(\frak{gl}(\mm|\nn))$ case as well. As before, the nonzero entries of 
the reflection matrix lie solely along the principal and the secondary diagonals of the bosonic
submatrices, or along 
the principal and one minor antidiagonal below or above the secondary one. 
For the symmetric Dynkin diagram, such nondiagonal solutions were first obtained in \cite{Doikou:2009dp}, 
albeit from a different point of view, where the nondiagonal entries lie only along the 
secondary diagonal. 

Substituting the ansatz (\ref{nond_struct_h}) into conditions (\ref{constr1})-(\ref{last_bosonic2}), we are 
able to obtain nondiagonal solutions for the distinguished Dynkin diagram that we consider in the present 
work. Note that the nonzero nondiagonal entries have either purely bosonic, or purely fermionic 
indices. For the distinguished grading, such solutions were first derived 
in \cite{Lima-Santos_slmn}\footnote{Note that these are purely bosonic 
reflection matrices. The author of \cite{Lima-Santos_slmn} did not consider
any fermionic degrees of freedom, so that Eq. (3.1) in \cite{Lima-Santos_slmn} leads necessarily to block-diagonal, bosonic matrices. 
Switching on the fermionic degrees of freedom implies that Eq. (3.1) holds
automatically since (in the notation of \cite{Lima-Santos_slmn}) 
$\b_{i,j} k_{i,j} = 0$ due to the nilpotency of Grassmann numbers. A more 
general constraint is actually found to hold for fermionic matrices below, see
Eqs. (\ref{fermi_diag_constraint}) and (\ref{Q_constr}) for the rational and
trigonometric cases respectively.}$^,$\footnote{A class of nondiagonal solutions of the ungraded model
was found in \cite{YangZhang}, corresponding to $\nn=0$ here.}. 
According to the terminology of \cite{Lima-Santos_slmn}, the solutions derived here correspond to 
solutions of Type II. In short, we have found the following families of solutions with four free
complex parameters $c_0, \ldots, c_3$, denote also $c_4 \equiv -c_0 + (c_0^2 + c_2 c_3)^{\frac{1}{2}}$:

\textbf{I$\ll$.} Solutions with bosonic nondiagonal indices for  $1\leq L \leq \frac{\mm}{2}$:
\small
\be
\begin{split}
& h_{\bar{\r}\r}(\l)=h_{\r\bar{\r}}(\l) = 0 \,,  \qquad 
h_{\r\r}(\l) = h_{\bar{\r}\bar{\r}}(\l) = c_0 + e^{-2\l} (c_1 + c_4 \sinh 2\l )\, , \\
& h_{jj} (\l)= c_0 + c_1\, e^{2\l}  , \qquad h_{\bar{j}\bar{j}}(\l) =c_0 + c_1\, e^{-2\l}, 
\cr
& h_{j\bar{j}}(\l) = c_2 \sinh2\l \, \qquad h_{\bar{j}j}(\l) = c_3 \sinh2\l \, , \quad 1\leq j\leq L \\
& h_{jj}(\l) = h_{\bar{j}\bar{j}}(\l) =  c_0 + c_1 \cosh 2\l + (c_1^2 + c_2 c_3)^{\frac{1}{2}} \sinh 2\l \, , \qquad 
h_{j\bar{j}} (\l)= h_{\bar{j}j} (\l)= 0 \, , \qquad L < j\leq \frac{\mm}{2}  \\
& h_{jj}(\l) = h_{\bar{j}\bar{j}}(\l) = c_0 + c_1 \cosh 2\l + (c_1^2 + c_2 c_3)^{\frac{1}{2}} \sinh 2\l 
\, , \qquad j = \bar{j} = \frac{\mm+1}{2} \, , 
\qquad \textrm{if} ~~ \mm ~~ \textrm{odd} \, .
\end{split}
\ee
\normalsize
\textbf{I$\gg$.}  Solutions with fermionic nondiagonal indices for $\mm+1 \leq \L \leq \mm + \frac{\nn}{2}$:
 \be
 \begin{split}
  & h_{j\bar{j}} (\l)= h_{\bar{j}j} (\l)= 0\, , \qquad 
  h_{jj}(\l) = h_{\bar{j}\bar{j}}(\l) = c_0+ e^{2\l}(c_1 + c_4 \sinh2\l) \, , \\
  & h_{\r\r} (\l)=c_0+c_1 \,e^{2\l} , \qquad h_{\bar{\r}\bar{\r}}(\l) = c_0+c_1 \,e^{-2\l}  , \cr
  & h_{\r\bar{\r}}(\l) = c_2 \sinh2\l \, , \qquad h_{\bar{\r}\r}(\l) = c_3 \sinh2\l , \quad \mm +1 \leq \r \leq \L\,, \\
  & h_{\r\r}(\l)= h_{\bar{\r}\bar{\r}}(\l) = c_0+c_1 \cosh 2\l + (c_1^2 + c_2 c_3)^{\frac{1}{2}} \sinh 2\l \, , \quad 
  h_{\bar{\r}\r}(\l)=h_{\r\bar{\r}}(\l) = 0 \, , \qquad \L < \r \leq \mm+\frac{\nn}{2} \\
  & h_{\r\r}(\l) = h_{\bar{\r}\bar{\r}}(\l) = c_0+c_1 \cosh 2\l + (c_1^2 + c_2 c_3)^{\frac{1}{2}} \sinh 2\l \, , \qquad 
  \r = \bar{\r} = \mm + \frac{\nn+1}{2} \, , \qquad \textrm{if} ~~ \nn ~~ \textrm{odd} \, , 
 \end{split}
 \ee
More reflection matrices with nondiagonal entries along minor antidiagonals below and above the secondary
one can be found in Appendix D.

\section{Fermionic reflection matrices}
Turning on the fermionic degrees of freedom implies the emergence of new
conditions that involve the anticommuting functions $\chi(\l)$. The conditions 
which contain only $h(\l)$ were written and partially solved in the previous section for 
various special cases, while the rest of them containing $\chi(\l)$ as well as those 
containing both $h(\l)$ and $\chi(\l)$ are collected 
in Appendix B. Focusing on the $\frak{gl}(\mm|\nn)$ symmetry and its $q$-deformation, it appears 
really hard, if not impossible, to solve relations 
(\ref{fermi_cond_first})-(\ref{fermi_cond_last}) for generic values of $\mm$ and $\nn$. For relatively 
small values of $\mm,\nn$ an analytical approach is possible, while for larger values one may resort
to an appropriate numerical analysis. For the purposes of the present work we consider the generic case and 
restrict to special solutions again by exploiting the results of the previous section for the bosonic 
parts of the solutions. 

\subsection{Solutions for the rational case with diagonal bosonic parts}
Recall that the diagonal bosonic solutions are given by the expressions
\be
K(\l) = \textrm{diag} \big( \underbrace{c_0 - \l , \cdots , c_0 - \l}_{\mathfrak{q}_1}, 
\underbrace{c_0 + \l , \cdots , c_0 + \l}_{\mathfrak{q}_2- \qf_1},
\underbrace{c_0 - \l , \cdots , c_0 - \l}_{\mm+\nn-\qf_2}, \big) \, ,
\label{K_diag_rational}
\ee
with $c_0$ being a free complex parameter and $0\leq \mathfrak{q}_1 \leq \qf_2 \leq \mm+\nn$.
Substituting these solutions first into relations (\ref{fermi_cond_6a})-(\ref{fermi_cond_last})
we find that the anticommuting functions have the following generic expressions
\be
\chi_{\g b}(\l) = \mathcal{C}_{\g b} \,  \mathcal{G}_{\g b} \, \l \,  W(\l) \, , \qquad \textrm{and} \qquad 
\chi_{c\s}(\l) = \mathcal{C}_{c\s} \,  \mathcal{H}_{c\s} \, \l \, W(\l) \, ,
\ee
with $\mathcal{C}_{\g b}, \mathcal{C}_{c\s}$ being sets of free complex parameters, 
$\mathcal{G}_{\g b}, \, \mathcal{H}_{c\s}$ denote sets of purely Grassmann numbers and $W(\l)$
is a function of the spectral parameter which depends on the values of $\qf_1,\qf_2$. Note that $W(\l)$ 
does not have an index dependence. However, it will be clear later that
not all of these complex and Grassmann parameters are nonzero or independent of each other.
It turns out that there
exist three families of solutions, which can be classified by the values of $\qf_1,\qf_2$
in (\ref{K_diag_rational}) as follows:
\paragraph{Class 1:} $(\qf_1,\qf_2) \in \{ (0,0), (0,\mm+\nn), (\mm,\mm+\nn)\}$. The reflection matrices 
that belong in this family have the maximal number of nonzero Grassmann 
variables. The lower left fermionic submatrix is full, while the upper right 
fermionic submatrix is full, but a row degeneracy exists: all entries of each row are 
equal. These statements are illustrated in the matrix below for $(\qf_1,\qf_2) = (\mm,\mm +\nn)$, 
where we have also defined 
the shorthands: $\mathcal{Q}_{\g b} \equiv \mathcal{C}_{\g b} \, \mathcal{G}_{\g b} \, \l$ and 
$\mathcal{Q}_{b} \equiv \mathcal{C}_{b \g} \, \mathcal{H}_{b\g} \, \l$
\be
K^{(1)}(\l) = 
\left(
\begin{array}{cccc|cccc}
 c_0 - \l & 0 & \cdots & 0 & 
 \mathcal{Q}_{1} \,  W_\l  &  \mathcal{Q}_{1} \, W_\l & \cdots & \mathcal{Q}_{1} \, W_\l \cr 
 0 & c_0 - \l & \ddots & \vdots  & \mathcal{Q}_{2}\, W_\l & \mathcal{Q}_{2} \, W_\l & \cdots 
 & \mathcal{Q}_{2} \, W_\l\cr 
 \vdots & \ddots & \ddots & 0 & \vdots & \vdots & \cdots & \vdots \cr 
 0 & \cdots & 0 & c_0 - \l &
  \mathcal{Q}_{\mm} \,  W_\l  &  \mathcal{Q}_{\mm} \,  W_\l & \cdots & \mathcal{Q}_{\mm} \, W_\l \cr \hline
 \mathcal{Q}_{\mm+1,1} \, W_\l  &  \mathcal{Q}_{\mm+1, 2}\,  W_\l 
 &  \cdots     &   \mathcal{Q}_{\mm+1,\mm}\,  W_\l & c_0 + \l & 0 & \cdots & 0 \cr
  \mathcal{Q}_{\mm+2,1} \, W_\l  &  \mathcal{Q}_{\mm+2, 2}  \,  W_\l  
  &  \cdots  &  \mathcal{Q}_{\mm+2,\mm} \, W_\l  & 0   &   c_0 + \l & \ddots & \vdots \cr
  \vdots  &   \vdots & \cdots &    \vdots  &  \vdots & \ddots & \ddots & 0 \cr 
   \mathcal{Q}_{\mm+\nn,1} \, W_\l  &  \mathcal{Q}_{\mm+\nn, 2}\,  W_\l  
   & \cdots  &   \mathcal{Q}_{\mm+\nn,\mm}\,  W_\l   & 0 &  \cdots  & 0 & c_0 + \l
\end{array}
\right) \, ,
\label{case1_rational}
\ee
The same structure holds for the cases $(\qf_1,\qf_2) = (0,0)$ and $(\qf_1,\qf_2)= (0, \mm+\nn)$ 
as well, with the function $W_\l\equiv W(\l)$ given by
\be
W(\l) = \left\{ 
\begin{array}{cc}
 \l-c_0\, , & \textrm{for}~(\qf_1,\qf_2) = (0,0) \cr
 1\, , & \textrm{for}~(\qf_1,\qf_2) = (\mm,\mm+\nn) \cr
 \l+c_0 \, , & \textrm{for}~(\qf_1,\qf_2) = (0,\mm+\nn) \, .
\end{array}
\right.
\ee
These solutions have in principle $\mm(\nn +1)$ free complex parameters and an equal number 
of free Grassmann numbers.

\paragraph{Class 2:} $\qf_1 \leq \qf_2 \leq \mm$ and $\qf_2 \neq 0$. 
The matrices that belong to this case can be considered as
special limits of the class 1 described above, since they have same expressions 
with (\ref{case1_rational}), but now $\mathcal{Q}_{i} = 0, \, \forall i\leq \qf_1$, $i>\qf_2$ and 
$\mathcal{Q}_{\g b}=0, \, \forall b \leq \qf_1, b>\qf_2,  \g \in[\mm+1,\mm+\nn]$. A representative
element of this class is written below 
\be
K^{(2)}_\l = \left(
\begin{array}{ccccccccc|ccc}
 \Xi^{-}_{\gg\gg} & 0 & \cdots & 0 & \cdots & 0 & 0 & \cdots & 0 &
 0  & \cdots & 0 \cr 
 0 & \ddots & \ddots & \vdots  & \cdots & \vdots & \vdots & \cdots & \vdots & \vdots & \cdots & \vdots \cr 
 \vdots & \ddots & \Xi^{-}_{\gg\gg} & 0 & \cdots & 0 & 0 & \cdots & 0 & 0 & \cdots & 0  \cr 
 0 & \cdots & 0 & \Xi^{+}_{\gg\gg} & \ddots & \ddots  & 0 & \cdots & 0 & 
  \mathcal{Q}_{\qf_1+1}   & \cdots & \mathcal{Q}_{\qf_1+1}  \cr
  \vdots & \cdots &  \vdots & \ddots & \ddots & \ddots & \vdots & \cdots & \vdots & \vdots  & \cdots & \vdots\cr 
  0 & \cdots & 0 & \ddots & \ddots & \Xi^{+}_{\gg\gg} & 0 & \cdots & 0 & 
  \mathcal{Q}_{\qf_2}   & \cdots & \mathcal{Q}_{\qf_2}  \cr 
  0 & \cdots & 0 & 0  & \cdots & 0 & \Xi^{-}_{\gg\gg} & \ddots & \vdots & 0 & \cdots & 0\cr 
  \vdots & \cdots & \vdots & \vdots & \cdots & \vdots  & \ddots & \ddots & 0 & \vdots & \cdots & \vdots \cr 
  0 & \cdots & 0 &  0 & \cdots  & 0  & \cdots  & 0 & \Xi^{-}_{\gg\gg} & 0 & \cdots & 0 \cr 
  \hline
 0  &  \cdots &  0 &  \mathcal{Q}_{\mm+1,\qf_1+1} & \cdots & \mathcal{Q}_{\mm+1,\qf_2}  
 & 0 & \cdots & 0 & \Xi^{-}_{\gg\gg} & \cdots & 0  \cr 
  \vdots  &   \cdots & \vdots &    \vdots  &  \cdots & \vdots & \vdots & \cdots & \vdots & \vdots &  \ddots & \vdots \cr 
   0  &  \cdots  & 0  &   
   \mathcal{Q}_{\mm+\nn,\qf_1+1} & \cdots & \mathcal{Q}_{\mm+\nn,\qf_2} & 0 & \cdots & 0 & 
   0 &  \cdots & \Xi^{-}_{\gg\gg}
\end{array}
\right) 
\ee
These solutions have in principle $(\qf_2-\qf_1)(\nn+1)$ free Grassmann parameters. 

\paragraph{Class 3:} $\mm < \qf_1 \leq \qf_2 \leq \mm+\nn$ or $\qf_1 \leq \mm < \qf_2$: The 
reflection matrices of this class have the minimal number of nonzero Grassmann boundary parameters,
since the upper right fermionic submatrix is always zero. A generic element of this class has the 
following expression for $\mm < \qf_1 \leq \qf_2 \leq \mm+\nn$
\be
K^{(3,1)}_\l= \left(
\begin{array}{ccc|cccccccccc}
 \Xi_{\gg\gg}^- & \cdots & 0 &  \cr 
 \vdots & \ddots & \vdots &   &   &  &  
 & \mathbb{O}_{\mm \times \nn} & &  &  &  \cr
 0 & \cdots & \Xi_{\gg\gg}^-  \cr  \hline
 0 & \cdots  &  0 & \Xi_{\gg\gg}^- &
  0  &  \cdots & 0 & \cdots & 0 & 0 & \cdots & 0 \cr 
 \vdots &  \cdots &  \vdots  &   0 & \ddots & \ddots & \vdots & \cdots & \vdots & \vdots & \cdots & \vdots \cr 
 0 & \cdots & 0 & \vdots & \ddots & \Xi_{\gg\gg}^- & 0 & \cdots & 0  & 0 & \cdots & 0 \cr
  \mathcal{Q}_{\qf_1+1,1}   &  \cdots & \mathcal{Q}_{\qf_1+1, \mm}   
  &  0  & \cdots   & 0 &  \Xi_{\gg\gg}^+ & \ddots & \ddots & 0 & \cdots & 0 \cr
  \vdots  &   \cdots & \vdots & \vdots  &  \cdots & \vdots & \ddots & \ddots & \ddots& \vdots & \cdots & \vdots \cr 
   \mathcal{Q}_{\qf_2,1}   &  \cdots
   & \mathcal{Q}_{\qf_2,\mm}   &    0  & \cdots &  0 & \ddots & \ddots & \Xi_{\gg\gg}^+ & 0 & \cdots & 0 \cr
   0  &  \cdots
   & 0 &    0  & \cdots &  0 & 0 & \cdots & 0 & \Xi_{\gg\gg}^- & \ddots  & \vdots \cr
   \vdots & \cdots & \vdots & \vdots & \cdots & \vdots & \vdots & \cdots & \vdots & \ddots & \ddots & 0 \cr
   0 & \cdots & 0 & 0 & \cdots & 0 & 0 & \cdots & 0 & \cdots & 0 & \Xi_{\gg\gg}^-
\end{array}
\right) 
\ee
and the initial number of free Grassmann parameters is $\mm(\qf_2-\qf_1)$.

For $\qf_1 \leq \mm < \qf_2$, a representative element of this case reads as  \small
\be
K^{(3,2)}_\l = \left(
\begin{array}{cccccc|ccccccc}
 \Xi_{\gg\gg}^- & 0 & \cdots  &  0 & \cdots & 0 \cr 
 0 & \ddots & \ddots &  \vdots  &  \cdots  & \vdots  &  
 & & &  &  &  \cr
 \vdots & \ddots & \Xi_{\gg\gg}^-  & 0 & \cdots & 0 \cr  
 0 & \cdots  &  0 & \Xi_{\gg\gg}^+ &
  \ddots  & \vdots & && \mathbb{O}_{\mm \times \nn} \cr 
 \vdots &  \cdots &  \vdots  &   \ddots & \ddots & 0 \cr 
 0 & \cdots & 0 & \cdots & 0 & \Xi_{\gg\gg}^+ \cr 
 \hline
  \mathcal{Q}_{\mm+1,1} &  \cdots & \mathcal{Q}_{\mm+1, \qf_1} 
  &  0  & \cdots   & 0 &  \Xi_{\gg\gg}^+ &  0 & \cdots & 0 & \cdots & 0 \cr
  \vdots  &   \cdots & \vdots &    \vdots  &  \cdots & \vdots & 0 & \ddots & \ddots& \vdots & \cdots & \vdots \cr 
   \mathcal{Q}_{\qf_2 ,1}  &  \cdots
   & \mathcal{Q}_{\qf_2, \qf_1}   & 0 & \cdots &  0 & \vdots & \ddots & \Xi_{\gg\gg}^+ & 0 & \cdots & 0 \cr
   0  &  \cdots & 0 &  
   \mathcal{Q}_{\qf_2+1, \qf_1+1}  & \cdots &  \mathcal{Q}_{\qf_2+1, \mm}  & 
   0 & \cdots & 0 & \Xi_{\gg\gg}^- & \ddots  & \vdots \cr
   \vdots & \cdots & \vdots & \vdots & \cdots & \vdots & \vdots & \cdots & \vdots & \ddots & \ddots & 0 \cr
   0 & \cdots & 0 & \mathcal{Q}_{\mm+\nn, \qf_1+1} & \cdots & \mathcal{Q}_{\mm+\nn, \mm} 
   & 0 & \cdots& 0 & \cdots & 0 & \Xi_{\gg\gg}^-
\end{array}
\right) 
\ee
\normalsize
with $\big[(\mm-\qf_1)(\mm+\nn-\qf_2)+\qf_1(\qf_2-\mm)\big]$  nonzero Grassmann parameters.

\subsubsection*{Constraints on the boundary parameters}
As stated before, in all of these three classes of reflection matrices not all of the complex or Grassmann
boundary parameters are nonzero or independent of each other. Proceeding to conditions 
(\ref{fermi_cond_first})-(\ref{fermi_cond_2c}), it is seen that they provide the constraints
\be
\mathcal{Q}_{c} \, \sum_\d  \mathcal{Q}_{\d b}  = 0 = \sum_d \mathcal{Q}_{d} \,  \mathcal{Q}_{\g d}   \, ,
\label{fermi_diag_constraint}
\ee
while conditions (\ref{B3a}), (\ref{B3b}) are automatically satisfied. the rest of the conditions 
are satisfied once the constraints (\ref{fermi_diag_constraint}) are taken into account. 
For generic values of $(\mm, \nn)$, one realizes that these constraints can be satisfied
in numerous ways. In particular, one can choose between setting some complex and Grassmann
boundary parameters to zero, and/or choosing specific Grassmann parameters that satisfy the 
constraint above, i.e. $\mathcal{H}_{c\r} \cdot \mathcal{G}_{\g b} = 0$. Therefore, the initial
number of free boundary parameters for the three aforementioned cases is reduced. It should 
also be stressed out that these constraints arose from the conditions containing solely fermionic
functions, and not the bosonic ones, thus they would be present in purely fermionic matrices 
as well. Taking into account the presentation of the three families
classified by the values of $\qf_1,\qf_2$, it should be apparent that when some of the $\mathcal{Q}$'s 
are already zero, as it happens in the classes 2 and 3, less constraints have to be chosen to 
satisfy (\ref{fermi_diag_constraint}).

\subsection{Solutions for the rational case with nondiagonal bosonic parts} 
Having classified the reflection matrices with diagonal bosonic parts, we continue our
analysis to matrices with nondiagonal bosonic parts, based on the findings of Section 3. We distinguish two 
classes of reflection matrices: 

\textbf{Class 1} with the bosonic parts given by $(\textbf{I}\ll, \textbf{II}\ll, \textbf{III}\ll$) and 

\textbf{Class 2} with the bosonic parts given by $(\textbf{I}\gg, \textbf{II}\gg, \textbf{III}\gg$). 

\no It turns out that the second class is just a special case of the first one. In the following, we 
describe the properties characterizing the structure of a generic element contained in  the first class:
\begin{enumerate}[(i)]
 \item Both fermionic submatrices are in general nonzero. The upper right fermionic submatrix in
 particular always possess some nonzero entries.
 \item As in the solutions containing diagonal bosonic submatrices, the upper right fermionic submatrix is 
 again row degenerate. 
 \item The entries of the upper and lower fermionic submatrices enjoy a \textit{mirror symmetry} with respect to the 
 $\left[\tfrac{\mm}{2}\right]$-th row or the $\left[\tfrac{\mm}{2}\right]$-th column respectively.
 \item If some of the bosonic entries of the secondary diagonal (or a minor one) are zero, for example
 the entries $h_{j_\ell\bar{j}_\ell}(\l) = h_{\bar{j}_\ell j_\ell}(\l) = 0$ for some $\ell$, then 
 the corresponding $j_\ell$-th and $\bar{j}_\ell$-th rows and columns of the upper fermionic and 
 lower fermionic matrices respectively are zero.
 \item For nondiagonal bosonic entries along a minor antidiagonal there are additional zero rows and 
 columns of the fermionic matrices, those corresponding to the positions of the entries $\Xi_{\gg\gg}^-(\l)$, 
 see also the exact structures $\textbf{II}\ll, \textbf{III}\ll$.
 \item The boundary parameters are again restricted through the constraint (\ref{fermi_diag_constraint}).
 \end{enumerate}

\no A representative element of \textbf{Class 1} is explicitly written in the next page. The reflection
matrices contained in $\textbf{Class 2}$ have similar structure to the one described above, whereas now
\begin{enumerate}[(a)]
 \item The upper right fermionic matrix is always zero.
 \item The entries of the upper and lower fermionic submatrices enjoy a \textit{mirror symmetry} with respect 
 to the ($\mm+\left[\tfrac{\nn}{2}\right]$)-th column or the ($\mm+\left[\tfrac{\nn}{2}\right]$)-th row respectively.
 \item The properties (iv), (v) and (vi) above are valid in this class as well.
\end{enumerate}


The following matrix is a representative element of the fermionic matrices contained in \textbf{Class 1}. 
The lower right bosonic submatrix is diagonal; the upper left one
has nonzero entries along the secondary diagonal apart from the entries 
$h_{\frac{\mm}{2}-1,\frac{\mm}{2}+1}(\l) = h_{\frac{\mm}{2}+1,\frac{\mm}{2}-1}(\l) = 0$. 
Without a loss of generality, it is assumed that $\mm$ is even. We also
denote $\mathcal{Q}_{\g b} \equiv \mathcal{C}_{\g b} \, \mathcal{G}_{\g b} \, \l$, ~
$\mathcal{Q}_{b} \equiv \mathcal{C}_{b\g} \, \mathcal{H}_{b\g} \, \l$, ~ 
$ \tilde{\mathcal{Q}}_{\g b} \equiv \tfrac{c_2}{c_4} \, \mathcal{Q}_{\g b}$ ~ and~
$\tilde{\mathcal{Q}}_{b} \equiv - \frac{c_1}{c_3} \, \mathcal{Q}_{b}$ \small

\begin{changemargin}{-1cm}{-1cm}

\[    
\left(
\begin{array}{cccccccccc|ccccc}
 \Xi_{\ll\ll}^+ & 0 & \cdots & \cdots & \cdots & \cdots & \cdots & \cdots & 0 & \Xi_{14}
 & \mathcal{Q}_{1 }  & \cdots & \mathcal{Q}_{1 }  \cr
 0 & \ddots & \ddots &&&&& \adots & \adots & 0 
 & \vdots & \cdots & \vdots \cr
 \vdots & \ddots & \Xi_{\ll\ll}^+ & \ddots &&& \adots & \Xi_{14} & \adots & \vdots 
 & \mathcal{Q}_{\frac{\mm}{2}-2}   & \cdots & \mathcal{Q}_{\frac{\mm}{2}-2}  \cr
  \vdots & & \ddots  & \Xi_{\gg\gg}^- & 0 & 0 &  0 & \adots && \vdots & 
   0 & \cdots & 0 \cr
  \vdots &&& 0 & \Xi_{\ll\ll}^+ & \Xi_{14} & 0 &&& \vdots 
  & \mathcal{Q}_{\frac{\mm}{2}}   & \cdots & \mathcal{Q}_{\frac{\mm}{2}}  \cr
  \vdots && & 0 & \Xi_{23} & \Xi_{\ll\ll}^- & 0 &&& \vdots & 
  \tilde{\mathcal{Q}}_{\frac{\mm}{2}}  & \cdots & \tilde{\mathcal{Q}}_{\frac{\mm}{2}}  \cr
  \vdots & & \adots & 0 & 0 & 0 & \Xi_{\gg\gg}^- & \ddots && \vdots 
  & 0 & \cdots &  0 \cr
  \vdots & \adots & \Xi_{23} & \adots &&& \ddots & \Xi_{\ll\ll}^- & \ddots & \vdots 
  & \tilde{\mathcal{Q}}_{\frac{\mm}{2}-2}  & \cdots & \tilde{\mathcal{Q}}_{\frac{\mm}{2}-2}  \cr
  0 & \adots & \adots &&&&& \ddots & \ddots & 0 
  & \vdots  & \cdots & \vdots \cr
  \Xi_{23} & 0 & \cdots & \cdots & \cdots & \cdots & \cdots & \cdots & 0 & \Xi_{\ll\ll}^-
  & \tilde{\mathcal{Q}}_{1} & \cdots & \tilde{\mathcal{Q}}_{1}  \cr 
  \hline
  \mathcal{Q}_{\mm+1,1} & \cdots & \mathcal{Q}_{\mm+1,\frac{\mm}{2}-2}  & 0 & \mathcal{Q}_{\mm+1,\frac{\mm}{2}}  
  & \tilde{\mathcal{Q}}_{\mm+1,\frac{\mm}{2}}  & 0 & \tilde{\mathcal{Q}}_{\mm+1,\frac{\mm}{2}-2}  & \cdots &
  \tilde{\mathcal{Q}}_{\mm+1,1}  & \Xi_{\gg\gg}^- & \cdots & 0 \cr 
  \vdots & \cdots & \vdots & \vdots & \vdots & \vdots & \vdots & \vdots & \cdots & \vdots & \vdots & \ddots & \vdots \cr
  \mathcal{Q}_{\mm+\nn,1} & \cdots & \mathcal{Q}_{\mm+\nn,\frac{\mm}{2}-2}  & 0 & \mathcal{Q}_{\mm+\nn,\frac{\mm}{2}}  &
  \tilde{\mathcal{Q}}_{\mm+\nn,\frac{\mm}{2}}  & 0 & \tilde{\mathcal{Q}}_{\mm+\nn,\frac{\mm}{2}-2} & 
  \cdots & \tilde{\mathcal{Q}}_{\mm+\nn,1}  &
   0 & \cdots & \Xi_{\gg\gg}^-
\end{array}
\right) 
\]
\end{changemargin}
\normalsize

\no The structure described above in properties (i)-(v) is manifest: the upper right fermionic submatrix is row 
degenerate and both fermionic submatrices possess the announced mirror symmetry. In this element we have 
chosen nondiagonal entries along the secondary diagonal with only two vanishing entries. It is easily seen that
the corresponding rows and columns of the fermionic matrices also vanish. As in the case of diagonal bosonic
parts, the boundary parameters are not free and satisfy the very same constraint, mainly
\be
\mathcal{Q}_{c} \, \sum_\d  \mathcal{Q}_{\d b}  = 0 = \sum_d \mathcal{Q}_{d} \,  \mathcal{Q}_{\g d} \, .
\ee

\subsection{Solutions for the trigonometric case with diagonal bosonic parts}
Recall that in the trigonometric case the diagonal bosonic solutions are given by the expressions
\be
 K(\l) = \textrm{diag} \big( \underbrace{c_0\, e^{\l} , \cdots , c_0 \, e^{\l}}_{\mathfrak{q}}, 
\underbrace{c_0 \, e^{-\l}, \cdots , c_0 \, e^{-\l}}_{\mm + \nn - \mathfrak{q}} \big) \, ,
\label{K_diag}
\ee
with $c_0$ being a free complex parameter and $0\leq \mathfrak{q} \leq \mm+\nn$. Similarly to 
the rational case, we find that the anticommuting functions have now the following generic expressions
\be
\chi_{\g b}(\l) = \mathcal{C}_{\g b} \,  \mathcal{G}_{\g b} \, \sinh(\l) \,  W(\l) \, , \qquad \textrm{and} \qquad 
\chi_{c\s}(\l) = \mathcal{C}_{c\s} \,  \mathcal{H}_{c\s} \, \sinh(\l) \, W(\l) \, ,
\ee
with $\mathcal{C}_{\g b}, \mathcal{C}_{c\s}$ being sets of free complex parameters, 
$\mathcal{G}_{\g b}, \, \mathcal{H}_{c\s}$ denote sets of purely Grassmann numbers and $W(\l)$
is a function of the spectral parameter which depends on the value of $\qf$. As before,
not all of these complex and Grassmann parameters are nonzero or independent of each other. 
The solutions can be classified again into three distinct families, labelled now by the value 
of $\mathfrak{q}$ in (\ref{K_diag}) as follows:

\paragraph{Class 1. $\qf \in \{0, \mm,\mm+\nn \}$:} This family contains the reflection matrices 
with the maximal number of nonzero Grassmann 
variables. The lower left fermionic submatrix is full, while the upper right 
fermionic submatrix is full and row degenerate. 
A representative element of this class with $\qf=\mm$ is written below, where we now define: 
$\mathcal{Q}_{\g b} \equiv \mathcal{C}_{\g b} \, \mathcal{G}_{\g b} \, \sinh\l$ and 
$\mathcal{Q}_{b} = \mathcal{C}_{b \g} \, \mathcal{H}_{b\g} \, \sinh\l$
\be
K^{(1)}(\l) = 
\left(
\begin{array}{cccc|cccc}
 c_0 \, e^\l & 0 & \cdots & 0 & 
 \mathcal{Q}_{1} \,  W_\l  &  \mathcal{Q}_{1} \, W_\l & \cdots & \mathcal{Q}_{1} \, W_\l \cr 
 0 & c_0 \, e^\l & \ddots & \vdots & \mathcal{Q}_{2}\, W_\l & \mathcal{Q}_{2} \, W_\l & \cdots 
 & \mathcal{Q}_{2} \, W_\l\cr 
 \vdots & \ddots & \ddots & 0 & \vdots & \vdots & \cdots & \vdots \cr 
 0 & \cdots & 0 & c_0 \, e^\l &
  \mathcal{Q}_{\mm} \,  W_\l  &  \mathcal{Q}_{\mm} \,  W_\l & \cdots & \mathcal{Q}_{\mm} \, W_\l \cr \hline
 \mathcal{Q}_{\mm+1,1} \, W_\l  &  \mathcal{Q}_{\mm+1, 2}\,  W_\l 
 &  \cdots     &   \mathcal{Q}_{\mm+1,\mm}\,  W_\l & c_0 \, e^{-\l} & 0 & \cdots & 0 \cr
  \mathcal{Q}_{\mm+2,1} \, W_\l  &  \mathcal{Q}_{\mm+2, 2}  \,  W_\l  
  &  \cdots  &  \mathcal{Q}_{\mm+2,\mm} \, W_\l  & 0   &   c_0 \, e^{-\l} & \ddots & \vdots \cr
  \vdots  &   \vdots & \cdots &    \vdots  &  \vdots & \ddots & \ddots & 0 \cr 
   \mathcal{Q}_{\mm+\nn,1} \, W_\l  &  \mathcal{Q}_{\mm+\nn, 2}\,  W_\l  
   & \cdots  &   \mathcal{Q}_{\mm+\nn,\mm}\,  W_\l   & 0 &  \cdots & 0 & c_0 \, e^{-\l} 
\end{array}
\right) \, ,
\label{case1}
\ee
The same structure holds for the cases $\qf = 0$ and $\qf = \mm+\nn$ 
as well, with the function $W_\l\equiv W(\l)$ given by
\be
W(\l) = \left\{ 
\begin{array}{cc}
 2 e^{\l} \cosh \l\, , & \textrm{for}~\qf = 0 \cr
 1\, ,& \textrm{for}~ \qf = \mm \cr
 2 e^{-\l} \cosh \l\, , & \qquad\textrm{for}~\qf  = \mm+\nn \, .
\end{array}
\right.
\ee
These solutions have in principle $\mm(\nn +1)$ free complex parameters and an equal number 
of free Grassmann numbers.

\paragraph{Class 2. $0 < \mathfrak{q} < \mm$:} The matrices contained in this class can be 
considered as special limits of the above class, since they have same expressions 
with (\ref{case1}), but now $\mathcal{Q}_{i} = 0, \, \forall i\geq \mm -\mathfrak{q}$ and 
$\mathcal{Q}_{\g b}=0, \, \forall b \geq \mm -\mathfrak{q}, \g \in[\mm+1,\mm+\nn]$. A representative
element of this class has the following expression:
\be
K^{(2)}(\l) = 
\left(
\begin{array}{cccccc|ccc}
 c_0 \, e^\l & 0 & \cdots & 0 & \cdots & 0 & 
 \mathcal{Q}_{1}  & \cdots & \mathcal{Q}_{1}  \cr 
 0 & \ddots & \ddots & \vdots  & \cdots & \vdots & \vdots & \cdots & \vdots 
 \cr 
 \vdots & \ddots & c_0 \, e^{\l} & 0 & \cdots & 0 & 
 \mathcal{Q}_{\qf}    & \cdots & \mathcal{Q}_{\qf}  \cr 
 0 & \cdots & 0 & c_0 \, e^{-\l} & \ddots & \vdots & 
  0   & \cdots & 0 \cr
  \vdots & \cdots &  \vdots & \ddots & \ddots & 0 & \vdots  & \cdots & \vdots\cr 
  0 & \cdots & 0 &  \cdots  & 0 & c_0 \, e^{-\l} & 0 & \cdots & 0\cr \hline
 \mathcal{Q}_{\mm+1,1}   &  \cdots &  \mathcal{Q}_{\mm+1,\qf}  &   0 & \cdots & 0
 & c_0 \, e^{-\l} & \cdots & 0  \cr 
  \vdots  &   \cdots & \vdots &    \vdots  &  \cdots & \vdots & \vdots &  \ddots & \vdots \cr 
   \mathcal{Q}_{\mm+\nn,1}   &  \cdots  & \mathcal{Q}_{\mm+\nn,\qf}  &   
   0 & \cdots & 0 & 0 &  \cdots & c_0 \, e^{-\l} 
\end{array}
\right) \, .
\ee
These solutions have in principle $\qf(\nn+1)$ free complex/Grassmann 
boundary parameters.

\paragraph{Class 3. $\mm < \qf < \mm + \nn$:} This class contains the solutions $K(\l)$ 
with the minimal number of nonzero
boundary parameters. The upper right fermionic submatrix vanishes completely, while in 
the lower left one the following elements are zero: $\mathcal{Q}_{\g b} = 0, \, \forall \g \leq \qf, \, 
1\leq b \leq \mm$
\be
K^{(3)}(\l) = 
\left(
\begin{array}{ccc|ccccccc}
 c_0 \, e^\l & \cdots & 0  \cr 
 \vdots & \ddots & \vdots & &  & & \mathbb{O}_{\mm \times \nn} & & \cr
 0 & \cdots & c_0 \, e^\l  \cr  \hline
 0 & \cdots  &  0 & c_0 \, e^\l &
  0  &  \cdots & 0 & \cdots & 0 \cr 
 \vdots &  \cdots &  \vdots  &   0 & \ddots & \ddots & \vdots & \cdots & \vdots \cr 
 0 & \cdots & 0 & \vdots & \ddots & c_0 \, e^{\l} & 0 & \cdots & 0  \cr
  \mathcal{Q}_{\qf+1,1}  &  \cdots & \mathcal{Q}_{\qf+1, \mm}   
  &  0  & \cdots   & 0 &  c_0 \, e^{-\l} & \ddots & \vdots \cr
  \vdots  &   \cdots & \vdots &    \vdots  &  \cdots & \vdots & \ddots & \ddots & 0 \cr 
   \mathcal{Q}_{\mm+\nn,1}   &  \cdots
   & \mathcal{Q}_{\mm+\nn,\mm}  &    0  & \cdots &  0 & \cdots & 0 & c_0 \, e^{-\l} 
\end{array}
\right) \, .
\ee
These solutions have initially $\mm(\mm + \nn - \qf)$ free Grassmann parameters. 

\paragraph{Restrictions on the boundary parameters:}

Similarly to the rational case, not all of the boundary parameters are nonzero or independent
of each other. Conditions (\ref{fermi_cond_first})-(\ref{fermi_cond_2c}) give rise to the familiar constraints
\[
\mathcal{Q}_{c} \, \sum_\d  \mathcal{Q}_{\d b}  = 0 = \sum_d \mathcal{Q}_{d} \,  \mathcal{Q}_{\g d}   \, ,
\]
while conditions (\ref{B3a}), (\ref{B3b}) are automatically satisfied. Manipulating the four 
remaining conditions, one arrives at an even stricter constraint, namely
\be
\mathcal{Q}_{c} \,  \mathcal{Q}_{\g b}  = 0 \, , \qquad \forall b,c,\g\, . 
\label{Q_constr}
\ee
We assume that this tighter restriction reflects the lesser freedom of the $q$-deformed symmetry
against the undeformed one. 

Considering some very simple limits to check our procedure, we made contact with previously obtained 
results. In particular, the restriction to a super spin chain with a $\mathfrak{gl}(1|1)$ symmetry
provides the results obtained in \cite{Grab_Frahm} for the 
free fermion model, while for the $\mathcal{U}_q(\mathfrak{gl}(1|1))$ case we rederive the expressions obtained 
in \cite{Karaiskos:2013oca} for the small polaron model. Note that in the latter case the Grassmann boundary 
parameters satisfy the condition $\a\cdot\b = 0$, which is a special case of (\ref{Q_constr}), which
holds for generic values of $\mm,\nn$. 

\subsubsection*{Solutions for the trigonometric case with nondiagonal bosonic parts}
For the trigonometric case we were not able to find fermionic reflection matrices with the nondiagonal
bosonic parts that we considered here, a fact which we assume is associated with the more restricted nature
of the deformed symmetry.

\section{Discussion}
In the present paper we have worked out the conditions satisfied by the entries of a
$K$-matrix that obeys the graded reflection algebra, for a large class of supersymmetric models 
and focused on the cases of a $\fr{gl}(\mm|\nn)$ symmetry and its $q$-deformation. We employed 
a useful notation which 
enabled us to readily distinguish between bosonic and fermionic degrees of freedom.
Regarding purely bosonic solutions, we have rederived previous results while also
extracted a new family of nondiagonal reflection matrices for the rational case. These solutions were 
later used as the bosonic parts of fuller reflection matrices, with fermionic degrees of freedom 
switched on.

The extracted conditions are in principle sufficient to derive all possible 
solutions of the reflection algebra associated with the $\fr{gl}(\mm|\nn)$ or 
$\mathcal{U}_q(\mathfrak{gl}(\mm|\nn))$ symmetries, for generic values of $\mm,\nn$. In practice, 
however, it is more realistic to consider either cases with small $\mm,\nn$ and obtain analytic results, 
or resort to numerical means for larger values. It would be also interesting to inspect if 
particular patterns govern the structures of the reflection matrices so that full $K$-matrices
can be constructed for arbitrary values of $\mm,\nn$. A systematic 
analysis starting from $\mm=\nn=1$ and building up may establish the emergence
of such patterns. 

Having acquired these families of fermionic reflection matrices, the next natural step
is to solve the corresponding boundary models. Previous experience 
\cite{Grab_Frahm, Karaiskos:2013oca} has demonstrated that the presence of 
supersymmetry assists towards obtaining the exact spectrum of the model. Another
direction is the construction of operatorial solutions to the reflection algebra, so that
dynamical boundary degrees of freedom may be introduced. Similar results are expected to 
arise for soliton nonpreserving boundary conditions as well \cite{Arnaudon:2003gj, Doikou:2000yw}. These issues 
will be addressed in future publications. 


\newpage 
\appendix

\section{Generalized conjugate indices}
In the current context, conjugate indices are systematically exploited to obtain 
bosonic solutions of the reflection equation, as presented in Section 3. Their 
construction is based on the fact that the bosonic entries of an $(\mm+\nn) 
\times(\mm+\nn)$ supermatrix form square submatrices of dimensions 
$\mm$ and $\nn$. The conjugate index 
\be
 \bar{a} = \mm + 1 - a \qquad \textrm{and} \qquad 
 \bar{\r} = 2\mm + \nn + 1 - \r \, ,
 \label{conj_ind_ap}
\ee
runs along the secondary diagonal of these submatrices. In order to systematically
construct solutions whose nonzero entries lie along minor antidiagonals, above and
below the secondary one, it
is possible to define ``generalized'' conjugate indices as
\be
\begin{split}
& a_\ell^- = \mm + 1 - \ell - a \, , \quad \textrm{for} \quad 1\leq a \leq 
\mm -\ell \quad \textrm{and} \quad \ell \in \big[0,\mm-1\big] \\
& a_\ell^+ = \mm + 1 + \ell - a \, , \quad \textrm{for} \quad \ell +1 \leq a \leq 
\mm  \quad \textrm{and} \quad \ell \in \big[0,\mm-1\big]
\end{split}
\ee
\be
\begin{split}
& \r_\xi^- = 2\mm + \nn +1 - \xi - \r  \, , \qq 
\textrm{for} \quad \mm+1 \leq \r \leq \mm+\nn -\xi  \quad 
\textrm{and} \quad \xi \in \big[0, \nn-1\big]  \\
& \r_\xi^+ = 2\mm + \nn +1 + \xi - \r  \, , \qq 
\textrm{for} \quad \mm+1 +\xi \leq \r \leq \mm+\nn  \quad 
\textrm{and} \quad \xi \in \big[0, \nn-1\big]  
\end{split}
\ee
It is clear that for $\ell = 0 = \xi$ the above definitions degenerate to 
eq. (\ref{conj_ind_ap}). These generalized conjugate indices span minor anti-diagonals 
above, or below of the secondary one. These statements are illustrated in the following 
schematic matrix.
\[
\left(
\begin{array}{ccccc|ccc}
a^-_\ell &a^-_\ell  &a^-_\ell  &a^-_\ell  & 
 \bullet &  & & \cr 
a^-_\ell  & a^-_\ell & a^-_\ell &\bullet  & a^+_\ell  & & & \cr 
 a^-_\ell & a^-_\ell & \bullet & a^+_\ell & a^+_\ell  & 
 & \mathbb{O}_{\mm \times \nn}& \cr 
 a^-_\ell & \bullet & a^+_\ell & a^+_\ell & a^+_\ell  &  & & \cr
 \bullet & a^+_\ell & a^+_\ell &  a^+_\ell & a^+_\ell & 
  & & \cr \hline
 & & & &  & \r^-_\xi & \r^-_\xi & \bullet \cr 
 & &  \mathbb{O}_{\nn \times \mm}  & &  & \r^-_\xi & \bullet & \r^+_\xi \cr 
 & & & &  & \bullet &\r^+_\xi & \r^+_\xi
 \end{array} \right)
\]
Any matrix can be expressed solely in terms of these conjugate indices. For example any purely 
bosonic reflection matrix can be written as
\[
 K(\l) = \sum_{a=1}^{\mm-\ell} \sum_{\ell=0}^{\mm-1} h^\l_{aa^-_\ell} \, e_{aa^-_\ell}
 + \sum_{a= \ell + 1}^{\mm} \sum_{\ell=0}^{\mm-1} h^\l_{aa^+_\ell} \, e_{aa^+_\ell}
 + \sum_{\r = \mm + 1}^{\mm + \nn -\xi} \sum_{\xi=0}^{\nn-1} h^\l_{\r\r^-_\xi} \, e_{\r\r^-_\xi}
 + \sum_{\r = \mm + 1 + \xi}^{\mm + \nn} \sum_{\xi=0}^{\nn-1} h^\l_{\r\r^+_\xi} \, e_{\r\r^+_\xi} \, .
\]

\section{Conditions for fermionic solutions}

$\ll\ll\ll\ll:$
\begin{subequations}
\begin{flalign}
& \sum_\d \Big( f_{cj}^- \big [ g_{j\d}^+ \, \chi_{c\d}^\l \, \chi_{\d i} ^\m  - 
\d_{ci}\, \d_{kj} \, g_{c\d}^+ \, \chi_{k\d}^\m \, \chi_{\d j}^\l  \big] + \d_{cj} \, \d_{kj \, }\, 
 g_{j\d}^+ \big( g_{jk}^- \, \chi_{k\d}^\l \, \chi_{\d i}^\m 
 - g_{ji}^- \, \chi_{k\d}^\m \, \chi_{\d i}^\l \big) \Big)= 0 &  \label{fermi_cond_first} \\ 
  &  \sum_\d  \Big( f_{cj}^- \big[ g_{c\d}^+ \, \chi_{k\d}^\m \, \chi_{\d j}^\l 
- \d_{kj} \, \d_{ci} \,  g_{j\d}^+ \, \chi_{c\d}^\l \, \chi_{\d i} ^\m \big]
 - \d_{cj} \, \d_{kj} 
 \, g_{j\d}^+ \big( g_{jk}^- \, \chi_{k\d}^\l \, \chi_{\d i}^\m 
 - g_{ji}^- \, \chi_{k\d}^\m \, \chi_{\d i}^\l \big) \Big)= 0  &  \\
 & \sum_\d \Big( g_{j\d}^+ \big( g_{jk}^- \, \chi_{k\d}^\l \, \chi_{\d i}^\m 
 - g_{ji}^- \, \chi_{k\d}^\m \, \chi_{\d i}^\l \big) 
 + f_{cj}^-  \big[ \d_{jk} \, \d_{cj} \, g_{j\d}^+ \, \chi_{c\d}^\l \, \chi_{\d i} ^\m 
 +\d_{ji} \, \d_{cj} \, g_{c\d}^+ \, \chi_{k\d}^\m \, \chi_{\d j}^\l \big] \Big)
 =0 & 
\end{flalign}
\end{subequations}
$\gg\gg\gg\gg:$
\begin{subequations}
\begin{flalign}
& \sum_d\Big( f_{\g\n}^- \big [ g_{\n d}^+ \, \chi_{\g d}^\l   \chi_{d \r} ^\m  - 
\d_{\g\r}\d_{\k \n} g_{\g d}^+ \, \chi_{\k d}^\m  \chi_{d \n}^\l  \big] + \d_{\g\n} \d_{\k \n} 
g_{\n d}^+ \big( g_{\n \k }^- \, \chi_{\k d}^\l  \chi_{d \r}^\m 
 - g_{\n\r}^- \, \chi_{\k d}^\m  \chi_{d \r}^\l \big) \Big) = 0 & \\ 
 & \sum_d \Big( f_{\g\n}^-  \big[ g_{\g d}^+ \, \chi_{\k d}^\m   \chi_{d \n}^\l 
- \d_{\k \n}   \d_{\g\r}   g_{\n d}^+ \, \chi_{\g d}^\l  \chi_{d \r} ^\m \big]
 - \d_{\g\n}   \d_{\k \n}  
  g_{\n d}^+ \big( g_{\n \k }^- \, \chi_{\k d}^\l   \chi_{d \r}^\m 
 - g_{\n\r}^- \, \chi_{\k d}^\m   \chi_{d \r}^\l \big) \Big) = 0  & \\
 & \sum_d \Big(g_{\n d}^+ \big( g_{\n \k }^- \, \chi_{\k d}^\l   \chi_{d \r}^\m 
 - g_{\n\r}^- \, \chi_{\k d}^\m   \chi_{d \r}^\l \big) 
 + f_{\g\n}^- \big[ \d_{\n \k }   \d_{\g\n}   g_{\n d}^+ \, \chi_{\g d}^\l   \chi_{d \r} ^\m 
 +\d_{\n\r}   \d_{\g\n}   g_{\g d}^+ \, \chi_{\k d}^\m   \chi_{d \n}^\l \big] \Big)
 =0  & \label{fermi_cond_2c}
\end{flalign}
\end{subequations}
\begin{subequations}
\begin{flalign}
& \ll\ll\gg\gg: \quad
f_{cj}^+  \big( g_{j\r}^- \, \chi_{\g j}^\m \, \chi_{c\r}^\l -
g_{c\g}^- \, \chi_{\g j}^\l \, \chi_{c\r}^\m \big)
+ \d_{cj} \, \sum_d g_{jd}^+ \big(g_{j\r}^- \, \chi_{\g d}^\m \, \chi_{d\r}^\l
- g_{j\g}^- \, \chi_{\g d}^\l \, \chi_{d\r}^\m  \big) =0 & \label{B3a}   \\
& \gg\gg\ll\ll: \quad 
f_{\g\s}^+ \big( g_{\s i}^- \, \chi_{c\s}^\m \, \chi_{\g i}^\l
- g_{\g c}^- \, \chi_{c\s}^\l \, \chi_{\g i}^\m \big)
+ \d_{\g\s} \sum_\d g_{\s\d}^+ \big(g_{\s i}^- \, \chi_{c\d}^\m \, \chi_{\d i}^\l
- g_{\s c}^- \, \chi_{c\d}^\l \, \chi_{\d i}^\m \big) =0 &
\label{B3b}
\end{flalign}
\end{subequations}
\begin{subequations}
\begin{flalign}
& \ll\gg\gg\ll: & \nonumber \\
& \big( f_{c\g}^- \, f_{\g\r}^+  -  f_{cj}^+ \, f_{j\r}^- \big) 
\chi_{c\r}^\l \, \chi_{\g j}^\m
+ \d_{cj} \,  f_{j\r}^- \sum_d g_{jd}^+ \, \chi_{\g d}^\m \, \chi_{d\r}^\l
+ \d_{\g\r} \, f_{c\r}^- \sum_\d g_{\r\d}^+ \, \chi_{c\d}^\l \, \chi_{\d j}^\m  =0 & \\
& \gg\ll\ll\gg: & \nonumber \\
& \big( f_{kj}^+ \, f_{\g k}^- - f_{\g\r}^+ \, f_{\r j}^-  \big) 
\chi_{\g j}^\l \, \chi_{k \r}^\m 
+ \d_{kj}\, f_{\g j}^-  \sum_d g_{jd}^+ \, \chi_{\g d}^\l\, \chi_{d\r}^\m 
+ \d_{\g\r} \, f_{\r j}^-  \sum_\d g_{\r\d}^+ \, \chi_{k\d}^\m \, \chi_{\d j}^\l  =0 &
\end{flalign}
\end{subequations}
\begin{subequations}
\begin{flalign}
&  \ll\gg\ll\gg :  \qquad  
f_{c\s}^+ \Big[ g_{ck}^- \, \chi_{k\s}^\l \, \chi_{c\r}^\m 
+ g_{\s\r}^- \, \chi_{k\s}^\m \, \chi_{c\r}^\l \Big]
+ \Big[ f_{ck}^- \, f_{k\s}^+ - f_{c\r}^+ \, f_{\r\s}^- \Big] \chi_{c\s}^\l \, \chi_{k\r}^\m = 0  & \\
& \gg\ll\gg\ll : \qquad 
f_{\g m}^+ \Big[ g_{\g\k}^- \, \chi_{\k m}^\l \, \chi_{\g j}^\m 
+ g_{mj}^- \, \chi_{\k m}^\m \, \chi_{\g j}^\l \Big]
+\Big[f_{\g\k}^- \, f_{\k m}^+ - f_{jm}^- \, f_{\g j}^+ \Big] \chi_{\g m}^\l \, \chi_{\k j}^\m =0 & 
\label{fermi_cond_5b}
\end{flalign}
\end{subequations}
\begin{subequations}
\begin{flalign}
 \gg\ll\gg\gg :  \qquad &
 f_{\g\k}^-\, f_{\k j}^+ \, h_{\k\r}^\m \, \chi_{\g j}^\l 
- f_{\g\r}^+ \, f_{\r m}^-\, h_{\k\r}^\m\, \chi_{\g j}^\l 
+ f_{\g j}^+ \big(g_{\g\k}^- \, h_{\g\r}^\m \,
\chi_{\k j}^\l - g_{j\r}^- \, h_{\g\r}^\l \, \chi_{\k j}^\m \big)  & \nonumber \\
& + \d_{\g\r} \, f_{\r j}^-  \big(\sum_d g_{\r d}^+ \, h_{dj}^\l \, \chi_{\k d}^\m 
-\sum_\d g_{\r\d}^+ \, h_{\k\d}^\m \, \chi_{\d j}^\l \big)  = 0 & \label{fermi_cond_6a}\\
 \ll\gg\ll\ll :  \qquad &
 \big(f_{ck}^- \, f_{k\s}^+ -f_{cj}(\l +\m) \, f_{j\s}^-  \big) h_{kj}^\m \, \chi_{c\s}^\l
+  f_{c\s}^+ \big( g_{ck}^- \, h_{cj}^\m \, \chi_{k\s}^\l 
- g_{\s j}^-\,  h_{cj}^\l \, \chi_{k\s}^\m \big) & \nonumber \\
& + \d_{jc} \, f_{c\s}^- \big(\sum_\d g_{c\d}^+ \, h_{\d\s}^\l \, \chi_{k\g}^\m 
-\sum_d g_{cd}^+ \, h_{kd}^\m\, \chi_{d\s}^\l \big) =0 & \\
 \ll\gg\gg\gg :  \qquad &
 f_{c\s}^+ \big( g_{c\g}^- \, h_{\g\s}^\l \, \chi_{c\r}^\m 
- g_{\s\r}^- \, h_{\g\s}^\m \, \chi_{c\r}^\l \big) 
+ \big( f_{c\g}^- \, f_{\g\s}^+ 
- f_{c\r}^+ \, f_{\r\s}^- \big) h_{\g\r}^\m \,  \chi_{c\s}^\l & \nonumber \\
&  + \d_{\s\g} \, f_{c\g}^- \big( \sum_\d g_{\g\d}^+ \, h_{\d\r}^\m \, \chi_{c\d}^\l
- \sum_d g_{\g d}^+ \, h_{cd}^\l \, \chi_{d\r}^\m \big)  = 0 & \\ 
\gg\ll\ll\ll : \qquad &
f_{\g m}^+ \big(g_{\g k}^- \, h_{km}^\l \, \chi_{\g j}^\m 
- g_{mj}^- \, h_{km}^\m \, \chi_{\g j}^\l \big)
+ \big( f_{km}^+ \, f_{\g k}^-  
- f_{jm}^- \, f_{\g j}^+ \big) h_{kj}^\m \, \chi_{\g m}^\l & \nonumber\\
& + \d_{km} \, f_{\g m}^- \big( \sum_d g_{md}^+ \, h_{dj}^\m \, \chi_{\g d}^\l 
- \sum_d g_{m\d}^+ \, h_{\g\d}^\l \, \chi_{\d j}^\m \big) = 0  & \\
 \ll\ll\ll\gg :  \qquad & 
 f_{cm}^+\big( g_{m\r}^- \, h_{jm}^\m \, \chi_{c\r}^\l 
+g_{cj}^- \, h_{jm}^\l \, \chi_{c\r}^\m \big) 
 + \big(f_{cj}^- \, f_{jm}^+ - f_{c\r}^+ \, f_{\r m}^- \big)h_{cm}^\l \, \chi_{j\r}^\m & \nonumber \\
& + \d_{jm} \, f_{cm}^- \big(\sum_d g_{md}^+ \, h_{cd}^\l \, \chi_{d\r}^\m 
-\sum_\d g_{m\d}^+ \, h_{\d\r}^\m \chi_{c\d}^\l \big) 
 + \d_{cm} \sum_d g_{md}^+ \big( g_{m\r}^- \, h_{jd}^\m \, \chi_{d\r}^\l
+ g_{mj}^- \, h_{jd}^\l \, \chi_{d\r}^\m  \big) & \nonumber \\
& - \d_{cm} \sum_\d g_{m\d}^+ 
\big(g_{mj}^- \, h_{\d\r}^\m \, \chi_{j\d}^\l 
+ g_{m\r}^- \, h_{\d\r}^\l \, \chi_{j\d}^\m \big) \big)  = 0 & \\
\gg\gg\ll\gg :  \qquad
&  \big(f_{k \s}^+ \, f_{\g k}^- - f_{\g\r}^+ \, f_{\r\s}^- \big) h_{\g\s}^\l
\, \chi_{k\r}^\m
- f_{\g\s}^+ \big(g_{\g k}^- \, h_{\g\r}^\m \, \chi_{k\s}^\l 
+ g_{\s\r}^- \, h_{\g\r}^\l \, \chi_{k \s}^\m\big) \nonumber \\
& +\d_{\g\r} \, f_{\g\s}^- \big(\sum_d g_{\g d}^+ \, h_{kd}^\m \, \chi_{d \s}^\l
- \sum_\d g_{\g\d}^+ \, h_{\d\s}^\l \, \chi_{k \d}^\m \big) 
+ \d_{\g\s}  \sum_d g_{\g d}^+ \big(g_{\g\r}^- \, h_{kd}^\m \, \chi_{d\r}^\l
 + g_{\g k}^- \, h_{kd}^\l \, \chi_{d\r}^\m \big) \nonumber \\
& - \d_{\g\s} \sum_\d g_{\g\d}^+ \big( g_{\g k}^- \, h_{\d\r}^\m \, \chi_{k\d}^\l
 + g_{\g\r}^- \, h_{\d\r}^\l \, \chi_{k \d}^\m \big)   =0 &\\
\ll\ll\gg\ll : \qquad
 & f_{cj}^\l \big(g_{c\g}^- \, h_{ci}^\m \, \chi_{\g j}^\l 
+ g_{ji}^- \, h_{ci}^\l \, \chi_{\g j}^\m \big)  
- \big(f_{ci}^+ \, f_{ij}^- - f_{c\g}^- \, f_{\g j}^+ \big) h_{cj}^\l \, \chi_{\g i}^\m
\nonumber \\
& - \d_{cj} \sum_d g_{jd}^+ \big(g_{j\g}^- \, h_{di}^\m \, \chi_{\g d}^\l 
+ g_{ji}^- \, h_{di}^\l \, \chi_{\g d}^\m \big) 
 + \d_{cj} \sum_\d g_{j\d}^+\big( g_{ji}^- \, h_{\g\d}^\m \, \chi_{\d i}^\l 
+ g_{j\g}^- \, h_{\g\d}^\l \, \chi_{\d i}^\m \big) \nonumber \\
& + \d_{ci} \, f_{cj}^- \big( -\sum_d g_{cd}^+ \, h_{dj}^\l \, \chi_{\g d}^\m 
+ \sum_\d g_{c\d}^+ \, h_{\g\d}^\m \, \chi_{\d j}^\l \big) =0 &\\
\gg\gg\gg\ll : \qquad
&  f_{\g\s}^+ \big(g_{\s j}^- \, h_{\r\s}^\m \, \chi_{\g j}^\l
+ g_{\g\r}^- \, h_{\r\s}^\l \, \chi_{\g j}^\m \big) 
+ \big(f_{\g\r}^- \, f_{\r\s}^+ -f_{j\s}^- \, f_{\g j}^+ \big) h_{\g\s}^\l \, \chi_{\r j}^\m
 \nonumber \\
& + \d_{\s\r} \, f_{\g\s}^- \big(\sum_\d g_{\s\d}^+ \, h_{\g\d}^\l \, \chi_{\d j}^\m 
- \sum_d g_{\s d}^+ \, h_{dj}^\m \, \chi_{\g d}^\l \big) 
+\d_{\g\s} \sum_\d g_{\s\d}^+ \big(g_{\s\r}^- \, h_{\r\d}^\l \, \chi_{\d j}^\m 
+ g_{\s j}^- \, h_{\r\d}^\m \, \chi_{\d j}^\l \big) \nonumber \\
& - \d_{\g\s} \sum_d g_{\s d}^+ \big( g_{\s\r}^- \, h_{dj}^\m \, \chi_{\r d}^\l 
+ g_{\s j}^- \, h_{dj}^\l \, \chi_{\r d}^\m \big) =0 \label{fermi_cond_last} &
 \end{flalign}
\end{subequations}

 \section{Explicit expressions for small values of $(\mm,\nn)$}
In order to make the general structure of the reflection matrices presented in the 
text clearer, we illustrate here a few reflection matrices for small values of 
$(\mm, \nn)$. We essentially follow the notation of the main text and classify 
these matrices with respect to $\qf_1, \qf_2$ and $\qf$ for the rational and 
trigonometric case respectively. The following expressions are to be complemented
by the constraints (\ref{fermi_diag_constraint}) and (\ref{Q_constr}) for the boundary
parameters.

\subsubsection*{Rational case with diagonal bosonic parts}

\paragraph{$\bullet~~(\mm,\nn)=(1,1)$}

\[
 \begin{split}
&  (\qf_1,\qf_2)= \left\{ 
\begin{array}{c}
 (0,0) \cr
 (1,1) 
\end{array}
\right. :
\hspace{1cm}
\left(
\begin{array}{c|c}
c_0  -\l  & \mathcal{C}_{1} \, \mathcal{G}_1 \,  \l (\l-c_0)
   \cr \hline
  \mathcal{C}_{21} \, \mathcal{H}_{21} \, \l (\l-c_0)  & c_0 -\l \cr 
 \end{array} \right) \cr
 &(\qf_1,\qf_2)=(0,2): \hspace{1cm}
\left(
\begin{array}{c|c}
c_0 + \l  & \mathcal{C}_{1} \, \mathcal{G}_1 \,  \l (\l+c_0)
   \cr \hline
  \mathcal{C}_{21} \, \mathcal{H}_{21} \, \l (\l+c_0)  & c_0  +\l \cr 
 \end{array} \right) \cr
 & (\qf_1,\qf_2)=(1,2): \hspace{1cm}
\left(
\begin{array}{c|c}
c_0 - \l  & \mathcal{C}_{1} \, \mathcal{G}_1 \, \l   \cr \hline
  \mathcal{C}_{21} \, \mathcal{H}_{21} \,  \l  & c_0 + \l \cr 
 \end{array} \right)
 \end{split}
\]

\paragraph{$\bullet~~(\mm,\nn)=(2,1)$}

\[
\begin{split}
&  (\qf_1,\qf_2)= \left\{ 
\begin{array}{c}
 (0,0) \cr
 (1,1) 
\end{array}
\right.: \hspace{1cm} \left(
\begin{array}{cc|c}
c_0 -\l & 0 & \mathcal{C}_1 \, \mathcal{G}_1 \,   \l (\l-c_0) \cr 
0 & c_0 -\l  & \mathcal{C}_2 \, \mathcal{G}_2 \,  \l (\l-c_0) \cr
\hline 
 \mathcal{C}_{31} \, \mathcal{H}_{31} \, \l (\l-c_0)  & 
 \mathcal{C}_{32} \, \mathcal{H}_{32} \, \l (\l-c_0) 
 & c_0 -\l 
 \end{array} \right) \cr
& (\qf_1,\qf_2)=(0,1) : \hspace{1cm} \left(
\begin{array}{cc|c}
c_0+ \l  & 0 & \mathcal{C}_1 \, \mathcal{G}_1 \,  \l \cr 
0 & c_0 -\l & 0 \cr
\hline 
\mathcal{C}_{31} \, \mathcal{H}_{31} \,  \l & 0 & c_0 -\l
 \end{array} \right) \cr
 & (\qf_1,\qf_2)=(0,2) : \hspace{1cm} \left(
\begin{array}{cc|c}
c_0 + \l & 0 & \mathcal{C}_1 \, \mathcal{G}_1 \, \l \cr 
0 & c_0+ \l & \mathcal{C}_2 \, \mathcal{G}_2 \,  \l \cr
\hline 
 \mathcal{C}_{31} \, \mathcal{H}_{31} \,  \l  & 
 \mathcal{C}_{32} \, \mathcal{H}_{32} \,  \l & c_0 -\l
 \end{array} \right) \cr
 & (\qf_1,\qf_2)=(0,3): \hspace{1cm} \left(
\begin{array}{cc|c}
c_0 + \l & 0 & \mathcal{C}_1 \, \mathcal{G}_1 \,  \l (\l+c_0) \cr 
0 & c_0 + \l & \mathcal{C}_2 \, \mathcal{G}_2 \,  \l (\l+c_0) \cr
\hline 
 \mathcal{C}_{31} \, \mathcal{H}_{31} \, \l (\l+c_0)  & 
 \mathcal{C}_{32} \, \mathcal{H}_{32} \, \l (\l+c_0)  & c_0 + \l
 \end{array} \right) \cr
  & (\qf_1,\qf_2)=(1,2): \hspace{1cm} \left(
\begin{array}{cc|c}
c_0 - \l & 0 & 0 \cr 
0 & c_0 + \l & \mathcal{C}_1 \, \mathcal{G}_1 \,  \l \cr
\hline 
 0  & 
 \mathcal{C}_{32} \, \mathcal{H}_{32} \, \l   & c_0 - \l
 \end{array} \right)
\end{split} 
\]

\small

\paragraph{$\bullet~~(\mm,\nn)=(2,2)$}
\[
\begin{split}
&  (\qf_1,\qf_2)= \left\{ 
\begin{array}{c}
 (0,0) \cr
 (1,1) \cr
 (2,2)
\end{array}
\right.: \hspace{.5cm} \left(
\begin{array}{cc|cc}
c_0 -\l & 0 & \mathcal{C}_1 \, \mathcal{G}_1 \,   \l (\l-c_0) & \mathcal{C}_1 \, \mathcal{G}_1 \,   \l (\l-c_0) \cr 
0 & c_0 -\l  & \mathcal{C}_2 \, \mathcal{G}_2 \,  \l (\l-c_0) & \mathcal{C}_2 \, \mathcal{G}_2 \,   \l (\l-c_0) \cr
\hline 
 \mathcal{C}_{31} \, \mathcal{H}_{31} \, \l (\l-c_0)  & 
 \mathcal{C}_{32} \, \mathcal{H}_{32} \, \l (\l-c_0) 
 & c_0 -\l & 0 \cr 
  \mathcal{C}_{41} \, \mathcal{H}_{41} \, \l (\l-c_0)  & 
 \mathcal{C}_{42} \, \mathcal{H}_{42} \, \l (\l-c_0) & 0 & c_0 - \l 
 \end{array} \right) \cr
 & (\qf_1,\qf_2)=(0,1): \hspace{1cm} 
 \left(
\begin{array}{cc|cc}
c_0 + \l & 0 & \mathcal{C}_1 \, \mathcal{G}_1 \,   \l  & \mathcal{C}_1 \, \mathcal{G}_1 \,   \l  \cr 
0 & c_0 -\l  & 0 & 0 \cr
\hline 
 \mathcal{C}_{31} \, \mathcal{H}_{31} \, \l   & 0 & c_0 -\l & 0 \cr 
  \mathcal{C}_{41} \, \mathcal{H}_{41} \, \l   & 0 & 0 & c_0 - \l 
 \end{array} \right) \cr
& (\qf_1,\qf_2)=(0,2): \hspace{1cm} 
  \left(
\begin{array}{cc|cc}
c_0 +\l & 0 & \mathcal{C}_1 \, \mathcal{G}_1 \,   \l & \mathcal{C}_1 \, \mathcal{G}_1 \,   \l  \cr 
0 & c_0 +\l  & \mathcal{C}_2 \, \mathcal{G}_2 \,  \l  & \mathcal{C}_2 \, \mathcal{G}_2 \,   \l  \cr
\hline 
 \mathcal{C}_{31} \, \mathcal{H}_{31} \, \l   & 
 \mathcal{C}_{32} \, \mathcal{H}_{32} \, \l  
 & c_0 -\l & 0 \cr 
  \mathcal{C}_{41} \, \mathcal{H}_{41} \, \l   & 
 \mathcal{C}_{42} \, \mathcal{H}_{42} \, \l  & 0 & c_0 - \l 
 \end{array} \right) \cr
& (\qf_1,\qf_2)=(0,3): \hspace{1cm} 
  \left(
\begin{array}{cc|cc}
c_0 +\l & 0 & 0 & \mathcal{C}_1 \, \mathcal{G}_1 \,   \l  \cr 
0 & c_0 +\l  & 0 & \mathcal{C}_2 \, \mathcal{G}_2 \,   \l  \cr
\hline 
0  & 0 & c_0 + \l & 0 \cr 
  \mathcal{C}_{41} \, \mathcal{H}_{41} \, \l   & 
 \mathcal{C}_{42} \, \mathcal{H}_{42} \, \l  & 0 & c_0 - \l 
 \end{array} \right) \cr
& (\qf_1,\qf_2)=(0,4): \hspace{1cm} 
  \left(
\begin{array}{cc|cc}
c_0 +\l & 0 & \mathcal{C}_1 \, \mathcal{G}_1 \,   \l (\l+c_0) & \mathcal{C}_1 \, \mathcal{G}_1 \,   \l (\l+c_0) \cr 
0 & c_0 +\l  & \mathcal{C}_2 \, \mathcal{G}_2 \,  \l (\l+c_0) & \mathcal{C}_2 \, \mathcal{G}_2 \,   \l (\l+c_0) \cr
\hline 
 \mathcal{C}_{31} \, \mathcal{H}_{31} \, \l (\l+c_0)  & 
 \mathcal{C}_{32} \, \mathcal{H}_{32} \, \l  (\l+c_0)
 & c_0 + \l & 0 \cr 
  \mathcal{C}_{41} \, \mathcal{H}_{41} \, \l (\l+c_0)  & 
 \mathcal{C}_{42} \, \mathcal{H}_{42} \, \l (\l+c_0)  & 0 & c_0 +\l 
 \end{array} \right) \cr
& (\qf_1,\qf_2)=(1,2): \hspace{1cm} 
  \left(
\begin{array}{cc|cc}
c_0 -\l & 0 & 0  & 0 \cr 
0 & c_0 +\l  & \mathcal{C}_2 \, \mathcal{G}_2 \,  \l  & \mathcal{C}_2 \, \mathcal{G}_2 \,   \l  \cr
\hline 
 0  & \mathcal{C}_{32} \, \mathcal{H}_{32} \, \l  
 & c_0 - \l & 0 \cr 
  0  & \mathcal{C}_{42} \, \mathcal{H}_{42} \, \l   & 0 & c_0 -\l 
 \end{array} \right) \cr
 & (\qf_1,\qf_2)=(1,3): \hspace{1cm} 
  \left(
\begin{array}{cc|cc}
c_0 -\l & 0 & \mathcal{C}_1 \, \mathcal{G}_1 \,   \l  & 0 \cr 
0 & c_0 +\l  & 0 & \mathcal{C}_2 \, \mathcal{G}_2 \,   \l \cr
\hline 
 \mathcal{C}_{31} \, \mathcal{H}_{31} \, \l  & 
 0  & c_0 + \l & 0 \cr 
  0 & \mathcal{C}_{42} \, \mathcal{H}_{42} \, \l   & 0 & c_0 - \l 
 \end{array} \right) \cr
 & (\qf_1,\qf_2)=(2,3): \hspace{1cm} 
  \left(
\begin{array}{cc|cc}
c_0 -\l & 0 & 0 & 0 \cr 
0 & c_0 -\l  & 0 & 0 \cr
\hline 
 \mathcal{C}_{31} \, \mathcal{H}_{31} \, \l  & 
 \mathcal{C}_{32} \, \mathcal{H}_{32} \, \l 
 & c_0 + \l & 0 \cr 
  0  & 0  & 0 & c_0 -\l 
 \end{array} \right) \cr
 & (\qf_1,\qf_2)=(3,4): \hspace{1cm} 
  \left(
\begin{array}{cc|cc}
c_0 -\l & 0 & 0 & 0 \cr 
0 & c_0 -\l  & 0 & 0 \cr
\hline 
0 & 0 & c_0 - \l & 0 \cr 
  \mathcal{C}_{41} \, \mathcal{H}_{41} \, \l  & 
 \mathcal{C}_{42} \, \mathcal{H}_{42} \, \l   & 0 & c_0 +\l 
 \end{array} \right)
 \end{split} 
 \]
\normalsize
\subsubsection*{Rational case with nondiagonal bosonic parts}
\paragraph{$\bullet~~(\mm,\nn)=(2,1):$}
\[ 
\left(
\begin{array}{cc|c}
c_0 +\frac{\l(c_1c_2+c_3c_4)}{c_3c_4-c_1c_2} & \frac{2\l c_2c_4}{c_1c_2-c_3c_4} & \mathcal{C}_1 \, \mathcal{G}_1 \,   \l \cr 
\frac{2\l c_2c_3}{c_3c_4-c_1c_2} & c_0 - \frac{\l(c_1c_2+c_3c_4)}{c_3c_4-c_1c_2}  & \mathcal{C}_2 \, \mathcal{G}_2 \,  \l  \cr
\hline 
 \mathcal{C}_{31} \, \mathcal{H}_{31} \, \l   & 
 \mathcal{C}_{32} \, \mathcal{H}_{32} \, \l 
 & c_0 -\l 
 \end{array} \right) 
\]
\paragraph{$\bullet~~(\mm,\nn)=(3,2):$}
\[ 
\left(
\begin{array}{ccc|cc}
c_0 +\frac{\l(c_1c_2+c_3c_4)}{c_3c_4-c_1c_2}  & 0  & \frac{2\l c_2c_4}{c_1c_2-c_3c_4} 
& \mathcal{C}_1 \, \mathcal{G}_1 \,   \l & \mathcal{C}_1 \, \mathcal{G}_1 \,   \l \cr 
0 & c_0 -\l  & 0
& 0 & 0 \cr 
\frac{2\l c_2c_3}{c_3c_4-c_1c_2} & 0 & c_0 - \frac{\l(c_1c_2+c_3c_4)}{c_3c_4-c_1c_2}  
& -\frac{c_1}{c_3} \, \mathcal{C}_1 \, \mathcal{G}_1 \,   \l & -\frac{c_1}{c_3} \, \mathcal{C}_1 \, \mathcal{G}_1 \,   \l \cr
\hline 
 \mathcal{C}_{41} \, \mathcal{H}_{41} \, \l   & 
 0 & \frac{c_2}{c_4} \, \mathcal{C}_{41} \, \mathcal{H}_{41} \, \l & c_0 -\l & 0 \cr
\mathcal{C}_{51} \, \mathcal{H}_{51} \, \l  &  0 & \frac{c_2}{c_4} \,\mathcal{C}_{51} \, \mathcal{H}_{51} \, \l
& 0 & c_0 -\l 
 \end{array} \right) 
\]
\paragraph{$\bullet~~(\mm,\nn)=(4,1):$}
\[ 
\left(
\begin{array}{cccc|c}
c_0 +\frac{\l(c_1c_2+c_3c_4)}{c_3c_4-c_1c_2}  & 0 & 0 & \frac{2\l c_2c_4}{c_1c_2-c_3c_4} 
& \mathcal{C}_1 \, \mathcal{G}_1 \,   \l \cr 
0 & c_0 - \l   & 0
& 0 &  0 \cr 
0 & 0 &  c_0 - \l
& 0 & 0 \cr
 \frac{2\l c_2c_3}{c_3c_4-c_1c_2}  & 
 0 & 0 & c_0 - \frac{\l(c_1c_2+c_3c_4)}{c_3c_4-c_1c_2} & -\frac{c_1}{c_3} \, \mathcal{C}_1 \, \mathcal{G}_1 \,   \l \cr
 \hline 
\mathcal{C}_{51} \, \mathcal{H}_{51} \, \l  &  0 & 0
& \frac{c_2}{c_4} \,\mathcal{C}_{51} \, \mathcal{H}_{51} \, \l & c_0 -\l 
 \end{array} \right) 
\]
\[ 
\left(
\begin{array}{cccc|c}
c_0 +\frac{\l(c_1c_2+c_3c_4)}{c_3c_4-c_1c_2}  & 0 & 0 & \frac{2\l c_2c_4}{c_1c_2-c_3c_4} 
& \mathcal{C}_1 \, \mathcal{G}_1 \,   \l \cr 
0 & c_0 +\frac{\l(c_1c_2+c_3c_4)}{c_3c_4-c_1c_2}  & \frac{2\l c_2c_4}{c_1c_2-c_3c_4} 
& 0 &   \mathcal{C}_2 \, \mathcal{G}_2 \,   \l \cr 
0 & \frac{2\l c_2c_3}{c_3c_4-c_1c_2} &  c_0 - \frac{\l(c_1c_2+c_3c_4)}{c_3c_4-c_1c_2}
& 0 & -\frac{c_1}{c_3} \, \mathcal{C}_2 \, \mathcal{G}_2 \,   \l \cr
 \frac{2\l c_2c_3}{c_3c_4-c_1c_2}  & 
 0 & 0 & c_0 - \frac{\l(c_1c_2+c_3c_4)}{c_3c_4-c_1c_2} & -\frac{c_1}{c_3} \, \mathcal{C}_1 \, \mathcal{G}_1 \,   \l \cr
 \hline 
\mathcal{C}_{51} \, \mathcal{H}_{51} \, \l  &  \mathcal{C}_{52} \, \mathcal{H}_{52} \, \l & 
\frac{c_2}{c_4} \,\mathcal{C}_{52} \, \mathcal{H}_{52} \, \l
& \frac{c_2}{c_4} \,\mathcal{C}_{51} \, \mathcal{H}_{51} \, \l & c_0 -\l 
 \end{array} \right) 
\]

 \subsubsection*{Trigonometric case with diagonal bosonic parts}
\paragraph{$\bullet~~(\mm,\nn)=(1,1)$}
\[
 \begin{split}
&  \qf=0: \hspace{1cm}
\left(
\begin{array}{c|c}
c_0 \, e^{-\l}  & \mathcal{C}_{1} \, \mathcal{G}_1 \,  e^\l \sinh2\l
   \cr \hline
  \mathcal{C}_{21} \, \mathcal{H}_{21} \, e^\l  \sinh2\l  & c_0 \, e^{-\l} \cr 
 \end{array} \right) \cr
 &\qf=1: \hspace{1cm}
\left(
\begin{array}{c|c}
c_0 \, e^\l  & \mathcal{C}_{1} \, \mathcal{G}_1 \,  \sinh\l
   \cr \hline
  \mathcal{C}_{21} \, \mathcal{H}_{21} \, \sinh\l  & c_0 \, e^{-\l} \cr 
 \end{array} \right) \cr
 & \qf=2: \hspace{1cm}
\left(
\begin{array}{c|c}
c_0 \, e^\l  & \mathcal{C}_{1} \, \mathcal{G}_1 \, e^{-\l}  \sinh2\l
   \cr \hline
  \mathcal{C}_{21} \, \mathcal{H}_{21} \, e^{-\l} \sinh2\l  & c_0 \, e^{\l} \cr 
 \end{array} \right)
 \end{split}
\]

\paragraph{$\bullet~~(\mm,\nn)=(2,1)$}

\[
\begin{split}
& \qf=0: \hspace{1cm} \left(
\begin{array}{cc|c}
c_0\, e^{-\l} & 0 & \mathcal{C}_1 \, \mathcal{G}_1 \,  e^\l \sinh2\l\cr 
0 & c_0\, e^{-\l} & \mathcal{C}_2 \, \mathcal{G}_2 \,  e^\l \sinh2\l\cr
\hline 
 \mathcal{C}_{31} \, \mathcal{H}_{31} \, e^\l  \sinh2\l  & 
 \mathcal{C}_{32} \, \mathcal{H}_{32} \, e^\l  \sinh2\l 
 & c_0\, e^{-\l}
 \end{array} \right) \cr
& \qf=1: \hspace{1cm} \left(
\begin{array}{cc|c}
c_0\, e^{\l} & 0 & \mathcal{C}_1 \, \mathcal{G}_1 \,   \sinh\l \cr 
0 & c_0\, e^{-\l} & 0 \cr
\hline 
\mathcal{C}_{31} \, \mathcal{H}_{31} \,  \sinh\l & 0 & c_0\, e^{-\l}
 \end{array} \right) \cr
 & \qf=2: \hspace{1cm} \left(
\begin{array}{cc|c}
c_0\, e^{\l} & 0 & \mathcal{C}_1 \, \mathcal{G}_1 \,   \sinh\l \cr 
0 & c_0\, e^{\l} & \mathcal{C}_2 \, \mathcal{G}_2 \,   \sinh\l \cr
\hline 
 \mathcal{C}_{31} \, \mathcal{H}_{31} \,  \sinh\l  & 
 \mathcal{C}_{32} \, \mathcal{H}_{32} \,  \sinh\l & c_0\, e^{-\l}
 \end{array} \right) \cr
 & \qf=3: \hspace{1cm} \left(
\begin{array}{cc|c}
c_0\, e^{\l} & 0 & \mathcal{C}_1 \, \mathcal{G}_1 \,  e^{-\l} \sinh2\l \cr 
0 & c_0\, e^{\l} & \mathcal{C}_2 \, \mathcal{G}_2 \,  e^{-\l} \sinh2\l \cr
\hline 
 \mathcal{C}_{31} \, \mathcal{H}_{31} \, e^{-\l}  \sinh2\l  & 
 \mathcal{C}_{32} \, \mathcal{H}_{32} \, e^{-\l}  \sinh2\l  & c_0\, e^{\l}
 \end{array} \right)
\end{split} 
\]

\paragraph{$\bullet~~(\mm,\nn)=(2,2)$}
\[
\begin{split}
& \qf=0: \hspace{1cm} \left(
\begin{array}{cc|cc}
c_0\, e^{-\l} & 0 & \mathcal{C}_1 \, \mathcal{G}_1 \,  e^\l \sinh2\l & 
\mathcal{C}_1 \, \mathcal{G}_1 \,  e^\l \sinh2\l \cr 
0 & c_0\, e^{-\l} & \mathcal{C}_2 \, \mathcal{G}_2 \,  e^\l \sinh2\l & 
\mathcal{C}_2 \, \mathcal{G}_2 \,  e^\l \sinh2\l \cr
\hline 
 \mathcal{C}_{31} \, \mathcal{H}_{31} \, e^\l  \sinh2\l  & 
 \mathcal{C}_{32} \, \mathcal{H}_{32} \, e^\l  \sinh2\l 
 & c_0\, e^{-\l} & 0 \cr
  \mathcal{C}_{41} \, \mathcal{H}_{41} \, e^\l  \sinh2\l  & 
 \mathcal{C}_{42} \, \mathcal{H}_{42} \, e^\l  \sinh2\l  & 0 & c_0\, e^{-\l}
 \end{array} \right) \cr
 & \qf=1: \hspace{1cm} \left(
\begin{array}{cc|cc}
c_0\, e^{\l} & 0 & \mathcal{C}_1 \, \mathcal{G}_1 \,   \sinh\l & 
\mathcal{C}_1 \, \mathcal{G}_1 \,  \sinh\l \cr 
0 & c_0\, e^{-\l} & 0 & 0 \cr
\hline 
 \mathcal{C}_{31} \, \mathcal{H}_{31} \, \sinh\l  & 
 0 & c_0\, e^{-\l} & 0 \cr
  \mathcal{C}_{41} \, \mathcal{H}_{41} \,   \sinh\l  & 
0  & 0 & c_0\, e^{-\l}
 \end{array} \right) \cr
 & \qf=2: \hspace{1cm} \left(
\begin{array}{cc|cc}
c_0\, e^{\l} & 0 & \mathcal{C}_1 \, \mathcal{G}_1 \,   \sinh\l & 
\mathcal{C}_1 \, \mathcal{G}_1 \,  \sinh\l \cr 
0 & c_0\, e^{\l} & \mathcal{C}_2 \, \mathcal{G}_2 \,  \sinh\l & 
\mathcal{C}_2 \, \mathcal{G}_2 \,   \sinh\l \cr
\hline 
 \mathcal{C}_{31} \, \mathcal{H}_{31} \,  \sinh\l  & 
 \mathcal{C}_{32} \, \mathcal{H}_{32} \,   \sinh\l 
 & c_0\, e^{-\l} & 0 \cr
  \mathcal{C}_{41} \, \mathcal{H}_{41} \,   \sinh\l  & 
 \mathcal{C}_{42} \, \mathcal{H}_{42} \,  \sinh\l  & 0 & c_0\, e^{-\l}
 \end{array} \right) \cr
 & \qf=3: \hspace{1cm} \left(
\begin{array}{cc|cc}
c_0\, e^{\l} & 0 & 0 & 0 \cr 
0 & c_0\, e^{\l} & 0 & 0 \cr
\hline 
 0  & 0 & c_0\, e^{\l} & 0 \cr
  \mathcal{C}_{41} \, \mathcal{H}_{41} \, e^\l  \sinh2\l  & 
 \mathcal{C}_{42} \, \mathcal{H}_{42} \, e^\l  \sinh2\l  & 0 & c_0\, e^{-\l}
 \end{array} \right) \cr 
 & \qf=4: \hspace{1cm} \left(
\begin{array}{cc|cc}
c_0\, e^{-\l} & 0 & \mathcal{C}_1 \, \mathcal{G}_1 \,  e^\l \sinh2\l & 
\mathcal{C}_1 \, \mathcal{G}_1 \,  e^\l \sinh2\l \cr 
0 & c_0\, e^{-\l} & \mathcal{C}_2 \, \mathcal{G}_2 \,  e^\l \sinh2\l & 
\mathcal{C}_2 \, \mathcal{G}_2 \,  e^\l \sinh2\l \cr
\hline 
 \mathcal{C}_{31} \, \mathcal{H}_{31} \, e^\l  \sinh2\l  & 
 \mathcal{C}_{32} \, \mathcal{H}_{32} \, e^\l  \sinh2\l 
 & c_0\, e^{-\l} & 0 \cr
  \mathcal{C}_{41} \, \mathcal{H}_{41} \, e^\l  \sinh2\l  & 
 \mathcal{C}_{42} \, \mathcal{H}_{42} \, e^\l  \sinh2\l  & 0 & c_0\, e^{-\l}
 \end{array} \right) 
\end{split}
\]

 \section{Bosonic nondiagonal solutions of the trigonometric case}

\textbf{II$\ll$.} Solutions with bosonic nondiagonal indices for  $1\leq L \leq \frac{\mm}{2} - 
\left[ \frac{\ell}{2}\right] $: \small
\be
\begin{split}
& h_{\r_\xi^\pm\r}(\l)=h_{\r\r_\xi^\pm}(\l) = 0 \,,  \qquad 
h_{\r\r}(\l) = h_{\r_\xi^\pm\r_\xi^\pm}(\l) = c_0 +  e^{-2\l}(c_1+c_4\sinh 2\l)\, , \\
& h_{jj} (\l)= c_0 + c_1\, e^{2\l}  , \qquad h_{j_\ell^-j_\ell^-}(\l) =c_0 + c_1\, e^{-2\l} \cr
& h_{jj_\ell^-}(\l) = c_2 \sinh2\l \, , \qquad h_{jj}(\l) = c_3 \sinh2\l \, , \qquad 1\leq j\leq L  \\
& h_{jj}(\l) = h_{j_\ell^-j_\ell^-}(\l) =  c_1 \cosh 2\l + (c_1^2 + c_2c_3 2)^{\frac{1}{2}} \sinh 2\l \, , \cr
& h_{jj_\ell^-} (\l)= h_{j_\ell^-j} (\l)= 0 \, , \qquad L < j \leq \frac{\mm}{2} - 
\left[ \frac{\ell}{2}\right] \\
& h_{jj} (\l) = h_{j_\ell^-j_\ell^-}(\l)  = c_0 + c_1 \cosh 2\l + (c_1^2 + c_2 c_3)^{\frac{1}{2}} \sinh 2\l \, , \quad 
j=j_\ell^- = \frac{\mm+1}{2}
-\left[\frac{\ell}{2}\right] \quad \textrm{if} ~~ \mm ~~\textrm{odd} \\
&  h_{jj}(\l) = c_0 +  e^{-2\l}(c_1+c_4\sinh 2\l) \, , \qquad \mm+1 - \ell \leq j \leq \mm \, .
\end{split}
\ee
\normalsize
\textbf{III$\ll$.} Solutions with bosonic nondiagonal indices for  $\ell +1 \leq L \leq \frac{\mm}{2} + 
\left[\frac{\ell}{2}\right]$: \small
\be
\begin{split}
& h_{\r_\xi^\pm\r}(\l)=h_{\r\r_\xi^\pm}(\l) = 0 \,,  \qquad 
h_{\r\r}(\l) = h_{\r_\xi^\pm\r_\xi^\pm}(\l) = c_0 + e^{-2\l}(c_1+c_4\sinh 2\l)\, , \\
& h_{jj}(\l) =  c_0 + e^{2\l}(c_1+c_4\sinh 2\l) \, , \qq 1 \leq j \leq \ell \\
& h_{jj} (\l)= c_0 + c_1\, e^{2\l}  , \qquad h_{j_\ell^+j_\ell^+}(\l) =c_0 + c_1\, e^{-2\l} \cr
& h_{jj_\ell^+}(\l) = c_2 \sinh2\l \, , \qquad h_{j_\ell^+j}(\l) = c_3 \sinh2\l \, , \qquad \ell  < j\leq L \\
& h_{jj}(\l) = h_{j_\ell^+j_\ell^+}(\l) =  c_0 +  c_1 \cosh 2\l + (c_1^2 + c_2c_3)^{\frac{1}{2}} \sinh 2\l \, , \cr
& h_{jj_\ell^+} (\l)= h_{j_\ell^+j} (\l)= 0 \, , \qquad L < j \leq  
\frac{\mm}{2} +  \left[\frac{\ell}{2}\right] \\
& h_{jj}(\l) = h_{j_\ell^+j_\ell^+}(\l) =  c_0 +  c_1 \cosh 2\l + (c_1^2 + c_2c_3)^{\frac{1}{2}} \sinh 2\l \, , \qquad 
j=j_\ell^+ = \frac{\mm+1}{2} + \left[\frac{\ell}{2}\right] \qquad
\textrm{if} ~~ \mm ~~ \textrm{odd}
\end{split}
\ee
\normalsize
\textbf{II$\gg$.} Solutions with fermionic nondiagonal indices for $\mm+1 \leq \L \leq \mm + \frac{\nn}{2} 
- \left[\frac{\xi}{2}\right]$: \small
\be
 \begin{split}
  & h_{jj_\ell^\pm} (\l)= h_{j_\ell^\pm j} (\l)= 0\, , \qquad 
  h_{jj}(\l) = h_{j_\ell^\pm j_\ell^\pm}(\l) = c_0 + e^{2\l}(c_1 + c_4 \sinh2\l) \, , \\
  & h_{\r\r} (\l)=c_1 \,e^{2\l} , \quad h_{\r_\xi^- \r_\xi^-}(\l) = c_1 \,e^{-2\l}  , \cr
  & h_{\r\r_\xi^-}(\l) = c_2 \sinh2\l , \qquad h_{\r_\xi^-\r}(\l) = c_3 \sinh2\l , \qquad \mm +1 \leq \r \leq \L \\
  & h_{\r\r}(\l)= h_{\r_\xi^-\r_\xi^-}(\l) = c_0 + c_1 \cosh 2\l + (c_1^2 + c_2 c_3)^{\frac{1}{2}} \sinh 2\l \, , \cr
  & h_{\r_\xi^-\r}(\l)=h_{\r\r_\xi^-}(\l) = 0 \, , \quad \L < \r \leq \mm+\frac{\nn}{2} - \left[\frac{\xi}{2}\right] \\
  & h_{\r\r}(\l)= h_{\r_\xi^-\r_\xi^-}(\l) = c_0 + c_1 \cosh 2\l + (c_1^2 + c_2 c_3)^{\frac{1}{2}} \sinh 2\l \, , \quad
  \r = \r_\xi^- = \mm + \frac{\nn+1}{2} - \left[\frac{\xi}{2}\right] \quad 
  \textrm{if}~~ \nn ~~ \textrm{odd} \\
  & h_{\r\r}(\l) = c_0 +  e^{-2\l}(c_1+c_4\sinh 2\l)  \, , \qquad  \mm + \nn + 1 - \xi \leq \r \leq \mm + \nn
 \end{split}
 \ee
\normalsize
 \textbf{III$\gg$.} Solutions with fermionic nondiagonal indices for $\mm+1 \leq \L \leq \mm + \frac{\nn}{2} 
+ \left[\frac{\xi}{2}\right]$ \small
\be
 \begin{split}
  & h_{jj_\ell^\pm} (\l)= h_{j_\ell^\pm j} (\l)= 0\, , \qquad 
  h_{jj}(\l) = h_{j_\ell^\pm j_\ell^\pm}(\l) = c_0 + e^{2\l}(c_1 + c_4 \sinh2\l) \, , \\
  & h_{\r\r}(\l) = c_0 + e^{2\l}(c_1+c_4\sinh 2\l) \, , \qquad \mm +1 \leq \r \leq \mm + \xi \\
  & h_{\r\r} (\l)=c_0 + c_1 \,e^{2\l} , \qquad h_{\r_\xi^+\r_\xi^+}(\l) = c_0 + c_1 \,e^{-2\l}  , \cr
  & h_{\r\r_\xi^+}(\l) = c_2 \sinh2\l , \qquad h_{\r_\xi^+\r}(\l) = c_3 \sinh2\l , \qquad \mm + \xi  < \r \leq \L\\
  & h_{\r\r}(\l)= h_{\r_\xi^+\r_\xi^+}(\l) = c_0 + c_1 \cosh 2\l + (c_1^2 + c_2c_3)^{\frac{1}{2}} \sinh 2\l \, , \cr
  & h_{\r\r_\xi^+}(\l)=h_{\r_\xi^+\r}(\l) = 0 \, , \qquad \L < \r \leq \mm+\frac{\nn}{2}  + \left[\frac{\xi}{2}\right] \\
  & h_{\r\r}(\l)= h_{\r_\xi^+\r_\xi^+}(\l) = c_0 + c_1 \cosh 2\l + (c_1^2 + c_2c_3)^{\frac{1}{2}} \sinh 2\l \, , \quad 
  \r = \r^+ = \m + \frac{\nn+1}{2} + \left[\frac{\xi}{2}\right] \quad 
  \textrm{if} ~~ \nn ~~ \textrm{odd}
 \end{split}
 \ee

\end{document}